\documentclass[11pt]{article}

\usepackage[margin=1in]{geometry}  
\usepackage{amsmath, amssymb, amsfonts}
\usepackage[cal=rsfs, scr=rsfs]{mathalfa}

\usepackage{amsthm}
\theoremstyle{plain}
\newtheorem{theorem}{Theorem}
\newtheorem{proposition}[theorem]{Proposition}

\theoremstyle{definition}
\newtheorem{definition}{Definition}

\theoremstyle{remark}

\usepackage{graphicx}
\usepackage[compatibility=false]{caption}
\usepackage{subcaption}
\usepackage{multirow}
\usepackage{longtable}
\usepackage{booktabs}

\usepackage{algorithm}
\usepackage{algorithmicx}
\usepackage{algpseudocode}

\usepackage{listings}
\usepackage{xcolor}
\usepackage{textcomp}
\usepackage{euscript}
\usepackage[title]{appendix}
\usepackage{hyperref}

\usepackage[numbers]{natbib}      
\title{A Stable and Theoretically Grounded Gromov-Wasserstein Distance for Reeb Graph Comparison using Persistence Images}
\author{
    Erin~W.~Chambers, and
    Guangyu~Meng \\
    Department of Computer Science and Engineering, University of Notre Dame, USA
}
\date{}

\begin{document}
\maketitle
\begin{abstract}
    Reeb graphs are a fundamental structure for analyzing the topological and geometric properties of scalar fields. Comparing Reeb graphs is crucial for advancing research in this domain, yet existing metrics are often computationally prohibitive or fail to capture essential topological features effectively. In this paper, we explore the application of the Gromov-Wasserstein distance, a versatile metric for comparing metric measure spaces, to Reeb graphs. We propose a framework integrating a symmetric variant of the Reeb radius for robust geometric comparison, and a novel probabilistic weighting scheme based on Persistence Images derived from extended persistence diagrams to effectively incorporate topological significance. A key contribution of this work is the rigorous theoretical proof of the stability of our proposed Reeb Gromov-Wasserstein distance with respect to perturbations in the underlying scalar fields. This ensures that small changes in the input data lead to small changes in the computed distance between Reeb graphs, a critical property for reliable analysis. We demonstrate the advantages of our approach, including its enhanced ability to capture topological features and its proven stability, through comparisons with other alternatives on several datasets, showcasing its practical utility and theoretical soundness.
\end{abstract}


\section{Introduction}

In computational topology and topological data analysis (TDA), Reeb graphs have emerged as powerful tools for analyzing scalar fields---fundamental for understanding structured data \cite{bollen2021reeb, wang2024measure}. Together with their discrete approximations, mapper graphs, they encode the changing component structure of level sets, thereby providing a compact yet expressive representation of the underlying topology. The versatility of Reeb graphs has spurred their widespread adoption in diverse applications, including shape analysis in computer graphics \cite{hilaga2001topology, escolano2013complexity}, feature extraction in scientific visualization \cite{tierny20063d, ge2011data}, and as topological summaries in deep learning \cite{kumari2020shapevis, curry2023topologically}.

Despite the utility of Reeb graphs in TDA, a significant challenge lies in effectively quantifying their similarity. Numerous metrics have been developed, broadly categorized into direct calculations and compound methods. Direct calculation techniques include the bottleneck distance \cite{cohen2009extending, bjerkevik2016stability}, interleaving distance \cite{de2016categorified}, functional distortion distance \cite{bauer2014measuring}, Reeb graph edit distance \cite{di2016edit}, and the universal distance \cite{bauer2021reeb}, each offering unique perspectives and properties \cite{bollen2021reeb}. Compound methods, such as Decorated Reeb Graphs (DRGs) \cite{curry2024stability} and Multi-Dimensional Reeb Graphs (MDRGs) \cite{ramamurthi2022topological}, aim to enhance Reeb graph comparisons by incorporating additional structures. DRGs, for instance, utilize barcode decorations with multiscale bottleneck and Gromov-Hausdorff distances \cite{memoli2011gromov} to provide robust multiscale topological summaries and improve sensitivity to local geometric variations. MDRGs, conversely, rely on bottleneck and Laplace-Beltrami-based extended bottleneck distances \cite{rosenberg1997laplacian}, demonstrating strengths in capturing richer geometric and spectral properties of the underlying data.

However, these existing methods face several limitations. The bottleneck distance, for example, can overlook structural context, as its value is determined by the single most prominent difference between persistence diagrams, thereby ignoring the collective information from all other topological features and their interrelationships. Other direct metrics are often computationally intractable  because they may require solving complex combinatorial optimization problems \cite{bollen2021reeb}.Compound methods frequently suffer from scalability issues due to the high memory and computational costs associated with added decorations or operators. Furthermore, hyperparameter tuning presents a significant hurdle for these compound metrics, as achieving an effective balance among their constituent components is inherently complex.

To address these limitations and offer a theoretically robust framework, we propose a novel approach for comparing Reeb graphs that leverages the Gromov-Wasserstein (GW) distance. The GW distance, a metric from optimal transport (OT) theory, is designed to compare metric measure spaces \cite{khamis2024scalable, montesuma2024recent, memoli2011gromov}. It is an appealing prospect as it captures both local and global structural differences by aligning intrinsic metric structures. Moreover, the GW distance provides node-to-node correspondences, enabling interpretable insights into graph similarities \cite{memoli2011gromov}. Note that our formulation, defined via the exact GW problem (which is NP-hard), is made computationally feasible by adopting a practical and standard approach in the field \cite{peyre2019computational, flamary2021pot}: leveraging efficient iterative solvers. These solvers are widely recognized for providing high-quality approximate solutions in polynomial time (with respect to the number of samples), with demonstrated acceptable approximation error for many applications \cite{peyre2019computational,solomon2016entropic}.

Our main contributions involve tailoring the GW distance framework specifically for Reeb graphs and establishing its theoretical underpinnings. First, to define the metric structure on the Reeb graph, we introduce a symmetric variant of the Reeb radius. This variant addresses the asymmetry of the traditional Reeb radius \cite{curry2024stability} and the noise sensitivity of alternatives like shortest path distances \cite{li2023comparing}, thereby providing a robust and valid distance metric between nodes. Second, for the measure component, we propose a Persistence Image (PI) \cite{adams2017persistence}-based probability measure. This approach transforms the extended persistence diagram \cite{cohen2009extending, agarwal2004extreme} associated with the scalar field into a Persistence Image \cite{adams2017persistence}, from which node weights are subsequently derived. Such a measure reflects the global topological context and significance of features, offering a more nuanced weighting than uniform or simpler heuristic distributions commonly employed \cite{curry2024stability, li2023comparing}.

A cornerstone of this work is the formal proof of stability for our proposed Reeb Gromov-Wasserstein ($RGW_p$) distance. We demonstrate that the $RGW_p$ distance between two Reeb graphs---equipped with our symmetric Reeb radius and PI-based measure---is bounded by the supremum norm difference between their generating scalar functions. This stability guarantee is crucial for the metric's reliability in practical applications where data can be noisy or subject to minor variations.

Through experiments on several benchmark datasets, we demonstrate the practical advantages of our $RGW_p$ distance, including its high accuracy, enhanced ability to capture topological features, and competitive computational efficiency. The combination of a robust geometric metric, a topologically informed measure, and proven stability establishes our framework as a compelling new tool for Reeb graph analysis, showcasing its practical utility and theoretical soundness. 

The remainder of this paper is organized as follows: Section 2 reviews background concepts. Section 3 details our methodology, covering the symmetric Reeb radius and the PI-based probability measure. Section 4 presents the stability analysis of our $RGW_p$ distance. Section 5 presents an example for the $RGW_p$ workflow. Section 6 describes our experimental setup and results. Finally, Section 7 concludes the paper and discusses future work.

\section{Background}
\label{sec:background} 

In this section, we provide a brief review of foundational concepts essential for understanding our work. These include scalar fields, Morse functions, Reeb graphs, various distance metrics applicable to Reeb graphs (such as the Reeb Radius, which forms a basis for our symmetric variant), the theory of extended persistence, and the construction of Persistence Images (PIs). For readers seeking a more comprehensive treatment of Reeb graphs and their associated metrics, we recommend the survey papers by Yan et al.~\cite{yan2021scalar} and Bollen et al.~\cite{bollen2021reeb}. For a thorough exploration of persistent homology, the texts by Dey and Wang~\cite{DeyWang2021} and Oudot~\cite{Oudot2015} are excellent resources. A summary of the mathematical notation used throughout this paper can be found in Appendix~\ref{app:notation_table}.

\subsection{Reeb Graphs}
\label{ssec:reeb_graphs_bg} 

A \textit{scalar field}, sometimes referred to as an $\mathbb{R}$-space, is formally a pair $(X,f)$, where $X$ denotes a topological space and $f: X \to \mathbb{R}$ is a continuous, real-valued function defined on $X$. The \textit{dimension} of such a scalar field corresponds to the dimension of its domain $X$. Conceptually, a scalar field provides a mathematical structure for tracking the behavior and evolution of real-valued functions over topological spaces.

\begin{figure}[!t]
\centering
\includegraphics[ width=1\textwidth]{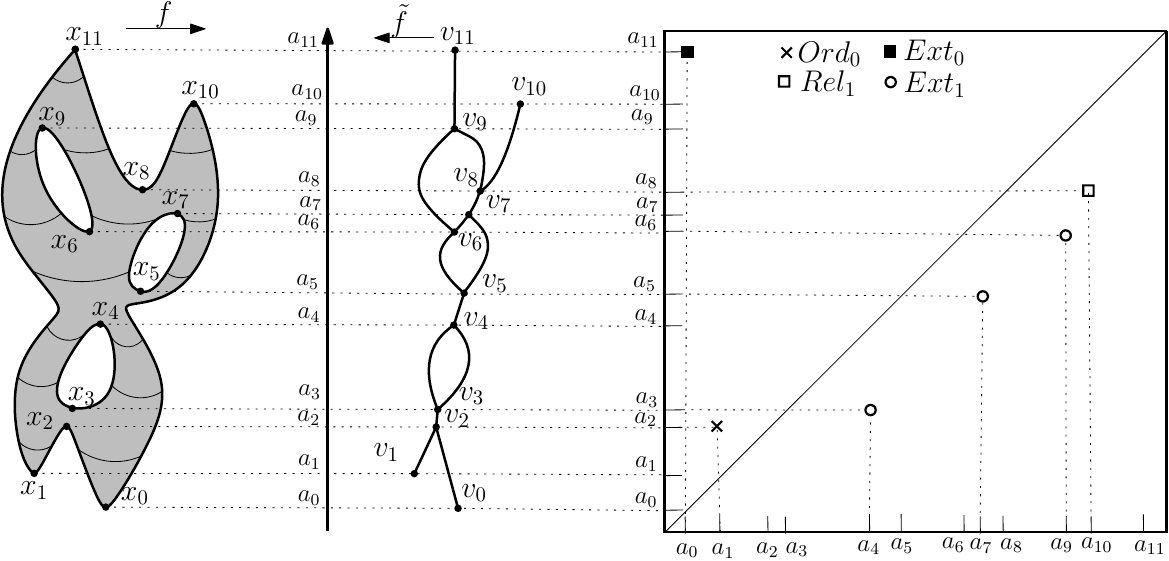}
\caption{(Left) A scalar field $(X,f)$, where $X$ is a compact 2-manifold and $f$ is a height function. The levels $a_0$ through $a_{11}$ represent critical values of the function $f$, and the points $x_0, x_1, \dots, x_{11}$ denote critical points on the manifold corresponding to these levels. (Middle) The Reeb graph $R_f$ of the scalar field $(X,f)$, where each node $v_0, v_1, \dots, v_{11}$ represents an equivalence class of points connected by level sets of $f$. (Right) The extended persistence diagrams for the scalar field.}
\label{fig:example}
\end{figure}

While scalar fields are pivotal for modeling data across numerous applications, the Reeb graphs derived from them may not possess desirable analytical properties unless certain constraints are imposed on the function $f$~\cite{de2016categorified}. A widely adopted and effective constraint is to require $f$ to be a \textit{Morse function}~\cite{milnor1963morse,yan2021scalar}. This ensures that the critical points of the scalar field are ``well-behaved," meaning they are non-degenerate.
More formally, a function $f: X \to \mathbb{R}$, where $X$ is a smooth manifold, is classified as a \textit{Morse function} if it is smooth and all of its critical points are \textit{non-degenerate}. A critical point $x \in X$ is termed non-degenerate if the Hessian matrix (the matrix of second partial derivatives) evaluated at $x$ is nonsingular. Furthermore, a Morse function is considered \textit{simple} if all of its critical points correspond to distinct function values.

For a scalar field $(X,f)$ where $f$ is a Morse function, and for any real number $a \in \mathbb{R}$, we define several key sets:
The \textit{level set} of $(X,f)$ at $a$ is the preimage $f^{-1}(a) := \{x \in X \mid f(x) = a\}$.
The \textit{sublevel set} of $(X, f)$ at $a$ is $f^{-1}((-\infty, a]) := \{x \in X \mid f(x) \leq a\}$.
The \textit{superlevel set} of $(X, f)$ at $a$ is $f^{-1}([a, \infty)) := \{x \in X \mid f(x) \geq a\}$.

The Reeb graph, illustrated in Figure~\ref{fig:example}, serves as a graph-based topological descriptor that concisely tracks how the topology of level sets evolves across the scalar field.
\begin{definition}[Reeb Graph] \label{def:reeb_graph}
Let \((X, f)\) be a scalar field. An equivalence relation \(\sim_f\) is defined on \(X\) such that \(x \sim_f y\) if and only if \(f(x) = f(y) = a\) for some \(a \in \mathbb{R}\), and both \(x\) and \(y\) reside in the same connected component of the level set \(f^{-1}(a)\).
The quotient space \(X / \sim_f\) is denoted by \(X_f\). For any equivalence class \([x] \in X_f\), we define a map $\bar{f}([x]) := f(x)$. Since \(x \sim_f y\) implies \(f(x) = f(y)\), this map \(\bar{f}: X_f \to \mathbb{R}\) is well-defined. We refer to \(\bar{f}\) as the \emph{map induced by \(f\) on the quotient}. The pair \(R_f := (X_f, \bar{f})\) is then defined as the \emph{Reeb graph} of \((X, f)\). The set of nodes (or vertices) of $R_f$ is denoted by $V_{R_f}$.
\end{definition}

\subsection{Distance Metrics on Reeb Graphs}
\label{sec:distance_metrics}
Several metrics can be employed to quantify the distance between nodes within the set $V_{R_f}$ of a Reeb graph $R_f$. 

\begin{definition}[Reeb Radius \cite{curry2024stability}] \label{def:reeb_radius}
For any two nodes $v, u \in V_{R_f}$, the \textit{Reeb radius} $\rho_f : V_{R_f} \times V_{R_f} \to \mathbb{R}_{\ge 0}$ is given by
\begin{equation}  
\rho_f(v, u) := \inf_{\gamma: v \to u} \sup_{t \in [0, 1]} |\bar{f}(v) - \bar{f}(\gamma(t))|,
\end{equation}

where the infimum is taken over all continuous paths $\gamma : [0,1] \to R_f$ within the Reeb graph that start at $v$ and end at $u$.
\end{definition}
For instance, in Figure~\ref{fig:example}, if we consider the induced map $\bar{f}$ on the Reeb graph nodes (which correspond to function values $a_i$), the Reeb radius $\rho_f(v_{11}, v_{10})$ would be $|\bar{f}(v_{11}) - \bar{f}(v_8)| = |a_{11} - a_8|$. It is important to note that $\rho_f$ is generally asymmetric, meaning $\rho_f(v, u) \neq \rho_f(u, v)$.

Beyond the Reeb radius, other metrics exist for comparing nodes within a Reeb graph. One such metric is the Reeb Distance.
\begin{definition}[Reeb Distance {\cite{bauer2014measuring}}] \label{def:reeb_distance}
Let $R_f$ be a Reeb graph with its induced map $\bar{f}$ and node set $V_{R_f}$. For any two nodes $u, v \in V_{R_f}$, the \textit{Reeb Distance}, denoted $\partial_f(u,v)$, is defined as:
\begin{equation} \label{eq:reeb_distance_definition}
\partial_f(u, v) := \inf_{\gamma \in \Gamma(u,v)} \mathrm{height}(\gamma).
\end{equation}
In this definition, $\Gamma(u,v)$ represents the set of all continuous paths $\gamma$ within the Reeb graph $R_f$ that connect node $u$ to node $v$. The height of any such path $\gamma$ is given by:
\begin{equation}
\mathrm{height}(\gamma) := \max_{x \in \gamma} \bar{f}(x) - \min_{x \in \gamma} \bar{f}(x).
\end{equation}
\end{definition}
For example, in Figure~\ref{fig:example}, the Reeb Distance $\partial_f(v_{11}, v_{10})$ is $|a_{11} - a_8|$.

The \textit{Shortest Path Distance}, denoted \( d_{sp} \), is defined as:
\begin{definition}[Shortest Path Distance] \label{def:shortest_path_dist_bg} 
\begin{equation}
d_{sp}(x, y) := \inf_{\gamma \in \Gamma(x, y)} \mathrm{length}(\gamma),
\end{equation}
where \(\Gamma(x, y)\) represents the set of all \emph{rectifiable} paths \(\gamma\) connecting nodes \(x, y \in V_{R_f}\). The Reeb graph is typically endowed with a metric graph structure where each edge, say between $u$ and $v$, has a length, commonly $|\bar{f}(u) - \bar{f}(v)|$. The length of a path $\gamma$, $\mathrm{length}(\gamma)$, is then the sum of the lengths of its constituent edges. Returning to Figure~\ref{fig:example}, the shortest path distance $d_{sp}(v_{11}, v_{10})$ would be calculated as $|a_{11} - a_9| + |a_9 - a_8| + |a_8 - a_{10}|$.
\end{definition}

\subsection{Extended Persistence}
\label{sec:extended_persistence}
Persistent homology stands as a fundamental tool in TDA, offering a robust methodology for studying the evolution of topological features—such as connected components, loops, and higher-dimensional voids—across varying scales. It meticulously tracks the ``birth" and ``death" of these features as a filtration parameter changes.

The primary output of this analysis is the \textit{persistence diagram}, denoted $D_f$. This diagram is a multiset of points \( (b, d) \in \mathbb{R}^2 \), where each point signifies a distinct topological feature that is born at scale (or function value) \( b \) and dies at scale \( d \). The \textit{persistence} of such a feature is then defined as $p = |d - b|$, representing its lifespan.

This paper specifically utilizes \textit{extended persistence diagrams}. These diagrams enrich the standard persistence framework by combining information from sublevel set homology (tracking features as function values increase from $-\infty$) with superlevel set relative homology (tracking features as function values decrease from $+\infty$, relative to a later-appearing feature). This combination provides a more complete topological summary of changes across the entire range of the scalar field, particularly as reflected in a Reeb graph~\cite{agarwal2004extreme,cohen2009extending}. The points $(b,d)$ in an extended persistence diagram, which correspond to critical values (e.g., $a_i, a_j$) and their associated Reeb graph vertices, are typically classified as follows:
\begin{itemize}
    \item \textbf{\(\mathbf{Ord_0}\) points $(a_i,a_j)$ (where $a_i<a_j$):} Represent a 0-dimensional feature (a connected component) that is born at $a_i$ and merges (dies) at $a_j$ during the sublevel set filtration.
    \item \textbf{\(\mathbf{Ext_0}\) points $(a_i,a_j)$ (where $a_i<a_j$):} Represent a 0-dimensional feature born in the sublevel set filtration at $a_i$ and persisting until it dies in the relative homology of a superlevel set filtration at $a_j$. These often correspond to global minima and maxima.
    \item \textbf{\(\mathbf{Rel_1}\) points $(a_j,a_i)$ (where $a_i<a_j$):} Represent a 1-dimensional feature (a loop) that is born in the sublevel set homology at $a_j$ and dies in the superlevel set relative homology at $a_i$.
    \item \textbf{\(\mathbf{Ext_1}\) points $(a_j,a_i)$ (where $a_i<a_j$):} These points typically pair an "upfork" (creation of multiple components from one) with a local maximum, appearing in the relative homology of superlevel sets.
\end{itemize}
Examples of these point types can be seen in Figure~\ref{fig:example} (right panel): $Ord_0 = \{(a_1, a_2)\}$, $Ext_0 = \{(a_0,a_{11})\}$, $Rel_1 = \{(a_{10}, a_8)\}$, and $Ext_1 = \{(a_4,a_3), (a_7,a_5), (a_9,a_6)\}$.

A Reeb graph is termed \textbf{generic} if it is derived from a simple Morse function defined on a compact manifold. This simplicity ensures that critical values are distinct, leading to unique persistence pairs associated with critical points. Our subsequent discussions will primarily focus on such generic Reeb graphs.

\subsection{Persistence Images}
\label{sec:persistence_images_bg}
Persistence Images (PIs) offer a method to transform persistence diagrams into a stable, finite-dimensional vector representation. This vectorization makes persistence diagrams more amenable to standard machine learning algorithms and other vector-based analytical techniques~\cite{adams2017persistence}. The process of constructing a Persistence Image \( I_f \in \mathbb{R}^N \) from an extended persistence diagram \(D_f = \{ (b_j, d_j) \}_j\) generally involves the following steps:

\begin{enumerate}
    \item \textbf{Coordinate Transformation}: Each birth-death pair \( (b_j, d_j) \in D_f \) is mapped to birth-persistence coordinates \( (b_j, p_j) \), where the persistence \( p_j = |d_j - b_j| \). This results in a new multiset of transformed points, \(D'_f = \{ (b_j, p_j) \}_j\). Points with zero persistence (i.e., $p_j=0$) are often disregarded or managed separately during this transformation.
    \item \textbf{Surface Generation}: A continuous surface \( \rho_{D'_f}(x, y) \) is generated in the birth-persistence plane. This is typically achieved by convolving the transformed points \( (b_j, p_j) \in D'_f \) with a kernel function, most commonly a Gaussian. Specifically, for each point \( (b_j, p_j) \), a weighted Gaussian kernel \( w(b_j, p_j) \, \phi_{(b_j, p_j)}(x, y) \) is centered at that point. The overall surface is the sum of these individual kernel contributions:
    \[
    \rho_{D'_f}(x, y) = \sum_{(b_j, p_j) \in D'_f} w(b_j, p_j) \, \phi_{(b_j, p_j)}(x, y).
    \]
    Here, \( w(b_j, p_j) \) is a non-negative weighting function, often chosen based on persistence (e.g., $w(b_j, p_j) = p_j$ to emphasize more persistent features, or $w(b_j,p_j)=1$ for uniform weighting). The term \( \phi_{(b_j, p_j)}(x, y) = \frac{1}{2\pi\sigma^2} \exp\left(-\frac{(x - b_j)^2 + (y - p_j)^2}{2\sigma^2}\right) \) represents a 2D Gaussian kernel with bandwidth (standard deviation) \( \sigma \).
    \item \textbf{Discretization}: The birth-persistence plane, or a specifically chosen region of interest within it, is then discretized into a grid of \( N \) pixels. The value of the \(k\)-th component of the Persistence Image vector \( I_f \), denoted \(I_f[k]\), is obtained by integrating the generated surface \( \rho_{D'_f}(x, y) \) over the area of the \(k\)-th pixel, \(P_k\):
    \[
    I_f[k] = \int_{P_k} \rho_{D'_f}(x, y) \, dx \, dy.
    \]
\end{enumerate}
The final result, \(I_f\), is a fixed-size vector, the Persistence Image. This stable vector representation is what we will leverage in our Methodology (Section~\ref{sec:methodology}) to define a probability measure on the nodes of a Reeb graph, thereby incorporating topological significance into our comparison framework.

\section{Methodology}
\label{sec:methodology}

In this section, we formally construct our proposed Gromov-Wasserstein distance tailored for Reeb graph comparison, which we denote as $RGW_p$. Our exposition will detail the two principal components that constitute this framework: first, the specific distance metric employed to quantify relationships between nodes within a Reeb graph—for which we use a Symmetric Reeb Radius; and second, the Persistence Image-based probability measure designed to assign topologically significant weights to these nodes.

\subsection{Gromov-Wasserstein Distance for Reeb Graphs}
\label{sec:rgw_definition}


Before specializing the GW distance to Reeb graphs, let us recall two foundational concepts:

\begin{definition}[Metric Measure Space (mm-space)]\label{def:mm_space_method} 
A \textit{metric measure space} (often abbreviated as mm-space) is a triple $(X, d_X, \mu_X)$, where $(X, d_X)$ constitutes a compact metric space (i.e., a set $X$ equipped with a distance function $d_X$ satisfying standard metric axioms), and $\mu_X$ is a Borel probability measure defined on $X$.
\end{definition}

\begin{definition}[Measure Coupling]\label{def:coupling_method} 
Given two metric measure spaces, $(X, d_X, \mu_X)$ and $(Y, d_Y, \mu_Y)$, a measure $\pi$ defined on the product space $X \times Y$ is termed a \textit{coupling} of $\mu_X$ and $\mu_Y$ if it satisfies the marginal conditions: for all measurable sets $A \subset X$ and $A' \subset Y$, we have $\pi(A \times Y) = \mu_X(A)$ and $\pi(X \times A') = \mu_Y(A')$. The set of all such valid couplings between $\mu_X$ and $\mu_Y$ is denoted by $\EuScript{M}(\mu_X, \mu_Y)$.
\end{definition}

We now adapt and specialize the Gromov-Wasserstein concept to the domain of Reeb graphs. Let $R_f^* = (R_f, d_{R_f}, \nu_{R_f})$ and $R_g^* = (R_g, d_{R_g}, \nu_{R_g})$ represent two Reeb graphs that have been decorated with metric measure spaces. Specifically:
\begin{itemize}
    \item $R_f$ and $R_g$ are the Reeb graphs derived from scalar fields $(X,f)$ and $(Y,g)$ respectively, with $V_{R_f}$ and $V_{R_g}$ denoting their respective sets of nodes.
    \item $d_{R_f}$ and $d_{R_g}$ are the distance metrics defined on the node sets $V_{R_f}$ and $V_{R_g}$, respectively. As we will elaborate in Section~\ref{sec:method_distance_metric}, we will employ the Symmetric Reeb Radius for this role.
    \item $\nu_{R_f}$ and $\nu_{R_g}$ are Borel probability measures defined on $V_{R_f}$ and $V_{R_g}$, respectively. Our choice for these measures, a PI-based approach, will be detailed in Section~\ref{sec:method_probability_measure}.
\end{itemize}

With these components, we define our Reeb Gromov-Wasserstein distance:

\begin{definition}[$RGW_p$ Distance for Reeb Graphs] \label{def:rgw_p_method}
The Reeb Gromov-Wasserstein distance ($RGW_p$) between two decorated Reeb graphs $R_f^*$ and $R_g^*$, for any $p \ge 1$, is defined as:
\[
RGW_p(R_f^*, R_g^*) := \inf_{\pi \in \EuScript{M}(\nu_{R_f}, \nu_{R_g})} \left( \sum_{v, v' \in V_{R_f}} \sum_{w, w' \in V_{R_g}} |d_{R_f}(v, v') - d_{R_g}(w, w')|^p \, \pi(v,w) \, \pi(v',w') \right)^{1/p}.
\]
Here, $\EuScript{M}(\nu_{R_f}, \nu_{R_g})$ is the set of all valid coupling measures (often termed transport plans) $\pi$ between the probability measures $\nu_{R_f}$ and $\nu_{R_g}$ defined on the node sets. In the discrete setting of Reeb graph nodes, $\nu_{R_f}$ and $\nu_{R_g}$ are probability vectors. The coupling $\pi$ becomes an $|V_{R_f}| \times |V_{R_g}|$ matrix, where each entry $\pi_{vw}$ quantifies the ``mass" transported from node $v \in V_{R_f}$ to node $w \in V_{R_g}$. This matrix must satisfy the marginal constraints: $\sum_{w \in V_{R_g}} \pi_{vw} = \nu_{R_f}(v)$ for each $v \in V_{R_f}$, and $\sum_{v \in V_{R_f}} \pi_{vw} = \nu_{R_g}(w)$ for each $w \in V_{R_g}$. The term $\pi(v,w) \pi(v',w')$ within the sum represents the joint probability of matching the pair of nodes $(v,v')$ from $R_f$ with the pair $(w,w')$ from $R_g$ according to the specific coupling plan $\pi$.
\end{definition}
This formulation effectively seeks an optimal correspondence between the node sets that minimizes the distortion of pairwise distances, weighted by the probability measures. The summation arises naturally from the integral definition of the Gromov-Wasserstein distance when applied to discrete spaces and measures.

It is important to note that while $RGW_p$ provides a robust measure of dissimilarity, the Gromov-Wasserstein framework, in general, does not guarantee that a distance of zero is obtained \textit{if and only if} the two metric measure spaces (and thus the Reeb graphs $R_f^*$ and $R_g^*$ in our context) are isomorphic. Specifically, two non-isomorphic metric measure spaces can be ``GW-isometric,'' resulting in a zero distance. Our stability results (Section 4) demonstrate continuity with respect to perturbations of the generating scalar fields for generic Reeb graphs, rather than a strict isomorphism-distinguishing property for all possible Reeb graph structures, including potentially degenerate cases like those that might arise from abstract graph embeddings.

\subsection{Distance Metric Component}
\label{sec:method_distance_metric}
\begin{figure}[bt]

\centering

\includegraphics[ width=0.5\textwidth]{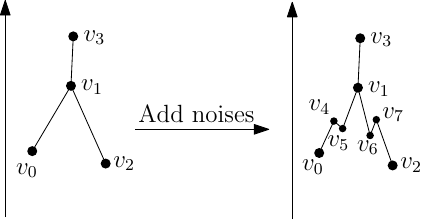}

\caption{An example illustrating that the shortest path distance is prone to noise. (Left) The original Reeb graph. (Right) A noisy version of the same Reeb graph.}
\label{fig:noise}

\end{figure}

A critical consideration in defining the $RGW_p$ distance is the choice of an appropriate intra-graph distance metric, $d_{R_f}$, for the nodes $V_{R_f}$ of a Reeb graph $R_f$. As briefly reviewed in Section~\ref{sec:distance_metrics}, several metrics have been proposed, each with distinct characteristics:
\begin{itemize}
    \item The standard \textbf{Reeb Radius} ($\rho_f$, Definition~\ref{def:reeb_radius}), while intuitively capturing functional variation along paths, suffers from asymmetry ($\rho_f(v, u) \neq \rho_f(u, v)$ in general). This makes it unsuitable as a direct metric component in the Gromov-Wasserstein framework, which inherently requires symmetric underlying metrics $d_X$ and $d_Y$.
    \item The \textbf{Reeb Distance} ($\partial_f$, Definition~\ref{def:reeb_distance}) is symmetric. However, it can violate the identity of indiscernibles, meaning distinct pairs of nodes might yield a zero Reeb distance if they happen to span the same range of function values along some connecting path.
    \item The \textbf{Shortest Path Distance} ($d_{sp}$, Definition~\ref{def:shortest_path_dist_bg}), though satisfying formal metric properties, often exhibits high sensitivity to noise and minor structural perturbations. As illustrated in Figure \ref{fig:noise}, even small perturbations in the graph structure can lead to significant variations in the shortest path, compromising its reliability in noisy settings. For instance, the original \( d_{sp}(v_0, v_2) \) is computed as \( |h(v_0) - h(v_1)| + |h(v_1) - h(v_2)| \), where \( h(v_i) \) denotes the height of each node. However, after introducing noise, the modified \( d_{sp}(v_0, v_2)' \) becomes $ |h(v_0) - h(v_4)| + |h(v_4) - h(v_5)| + |h(v_5) - h(v_1)| + |h(v_1) - h(v_6)| + |h(v_6) - h(v_7)| + |h(v_7) - h(v_2)|$, highlighting its instability. In contrast, both the Reeb radius and Reeb distance between \( (v_0, v_2) \) remain consistent even when noise is introduced.
\end{itemize}

To navigate these limitations, we adopt the \textbf{Symmetric Reeb Radius} as our chosen distance metric $d_{R_f}$ (and analogously $d_{R_g}$) for the $RGW_p$ computation.
\begin{definition}[Symmetric Reeb Radius] \label{def:symmetric_reeb_radius}
To ensure a valid and robust metric for comparing nodes within a Reeb graph $R_f$, the Symmetric Reeb Radius $d_{R_f}: V_{R_f} \times V_{R_f} \to \mathbb{R}_{\ge 0}$ is defined as:
\[
d_{R_f}(v, u) := \frac{1}{2} \left( \rho_f(v, u) + \rho_f(u, v) \right).
\]
This metric inherently satisfies non-negativity and symmetry. For generic Reeb graphs, it also satisfies the identity of indiscernibles and, crucially, the triangle inequality (see Proposition~\ref{prop:metric}), thus qualifying $(V_{R_f}, d_{R_f})$ as a metric space.
\end{definition}
By averaging the forward and backward Reeb radii, $d_{R_f}$ achieves the necessary symmetry for the GW framework while preserving the Reeb radius's core strength: its ability to capture functional variation along paths within the graph structure. This symmetrization yields a metric that is robust to noise than the shortest path distance and, unlike the Reeb distance, properly distinguishes between distinct nodes. The satisfaction of quasi-metric properties, formally stated in Proposition~\ref{prop:metric} and proven in Appendix~\ref{appendix:proof}, confirms its validity for use within the Gromov-Wasserstein distance. The stability of this specific metric choice will be a key focus of our analysis in Section~\ref{sec:stability_analysis}.

\begin{proposition}
The Symmetric Reeb Radius \( d_{R_f} \), as defined in Definition~\ref{def:symmetric_reeb_radius}, satisfies the properties of a metric (non-negativity, identity of indiscernibles for generic Reeb graphs, symmetry, and triangle inequality). It is therefore a quasi-metric suitable for the Gromov-Wasserstein framework.
\label{prop:metric}
\end{proposition}
The proof of this proposition is provided in Appendix~\ref{appendix:proof}.

\subsection{Probability Measure Component}
\label{sec:method_probability_measure}

The second crucial component in equipping a Reeb graph for the $RGW_p$ distance is the assignment of a Borel probability measure (e.g., $\nu_{R_f}$ on $V_{R_f}$ and $\nu_{R_g}$ on $V_{R_g}$). This measure is intended to reflect the relative importance or topological significance of different nodes within each Reeb graph. Several strategies for defining such measures exist:
\begin{itemize}
    \item \textbf{Uniform Distribution:} This straightforward approach assigns an equal weight to every node in the Reeb graph (e.g., as explored in \cite{li2023flexible, curry2023topologically}). While simple to implement, it inherently disregards the often substantial variations in topological importance among different nodes.
    \item \textbf{Intensity-Based Distribution:} In this strategy, nodes are weighted based on their associated scalar field values or local data densities (see, e.g., \cite{solomon2016entropic, peyre2019computational}). However, such measures can be unduly biased towards nodes with extreme function values and may inadvertently neglect topologically significant features (like saddles or short-lived components) that occur at moderate function values.
    \item \textbf{Degree-Based Distribution:} This method assigns weights to nodes according to their connectivity or degree within the graph structure (e.g., \cite{xu2019gromov, li2023comparing}). This approach is often unsuitable for generic Reeb graphs, where node degrees are typically very restricted (commonly 1 for leaves or 3 for saddles), rendering node degree an insufficiently discriminative proxy for genuine importance.
 \item \textbf{Lifespan-Based Distribution:} This more topologically informed heuristic utilizes the lifespan of persistence pairs to assign weights to Reeb graph nodes. The process generally involves:
        \begin{enumerate}
            \item Identifying features in the extended persistence diagram (Section~\ref{sec:extended_persistence}). Each feature, represented by critical values $(a_{\text{birth}}, a_{\text{death}})$, corresponds to specific nodes $v_{\text{birth}}$ and $v_{\text{death}}$ in the Reeb graph such that $\bar{f}(v_{\text{birth}}) = a_{\text{birth}}$ and $\bar{f}(v_{\text{death}}) = a_{\text{death}}$.
            \item Calculating the lifespan $p = |a_{\text{death}} - a_{\text{birth}}|$ for each feature. This lifespan $p$ is then attributed as a score to the associated nodes (e.g., both $v_{\text{birth}}$ and $v_{\text{death}}$ receive this score $p$). 
            \item Normalizing these accumulated scores for each node $v_i$, denoted $\text{score}(v_i)$, to obtain a probability distribution.
        \end{enumerate}

    While intuitively appealing for capturing feature significance, such direct lifespan-based measures exhibit notable drawbacks. They can suffer from \textit{instability}; minor perturbations in the underlying scalar field can not only shift the birth/death values of persistence features but can sometimes even alter the fundamental pairings of critical values that define these features. Either type of change can, in turn, lead to disproportionate variations in the normalized probability distribution assigned to nodes, a problem exacerbated when some lifespans are inherently small or when the total sum of lifespans (often used for normalization) itself changes significantly.  
\end{itemize}

To address these challenges and to develop a more robust and stable probability measure, we propose a novel approach that leverages \textbf{Persistence Images (PIs)}, as defined in Section~\ref{sec:persistence_images_bg}. The Persistence Image $I_f$ offers a stable, vector-based summarization of the entire extended persistence diagram $D_f$. From this PI, we derive the probability $\nu_{R_f}(v)$ for each relevant node $v \in V_{R_f}$ as follows:

\begin{definition}[PI-Based Probability Measure for Reeb Graph Nodes] \label{def:pi_based_measure_method}
Let $R_f$ be a Reeb graph, $D_f$ its associated extended persistence diagram, and $I_f$ the corresponding Persistence Image, constructed as detailed in Section~\ref{sec:persistence_images_bg}. For each node $v \in V_{R_f}$ that corresponds to an endpoint of a persistence interval (and thus has an associated birth-persistence point $(b_v, p_v)$ derived from $D'_f$):
\begin{enumerate}
    \item \textbf{Node Contribution from PI}: The contribution of a node $v$, denoted $\text{contrib}(v)$, is quantified by assessing how its specific persistence characteristics $(b_v, p_v)$ align with the global structure of the Persistence Image $I_f$. This is calculated as:
    \[
    \text{contrib}(v) := \sum_{k=1}^N I_f[k] \cdot \phi_{(b_v, p_v)}(c_k),
    \]
    where $I_f[k]$ is the value (integrated density) of the $k$-th pixel of the PI, $c_k$ is the geometric center of that $k$-th pixel, and $\phi_{(b_v, p_v)}(\cdot)$ is the same Gaussian kernel (with bandwidth $\sigma$, centered at the node's own birth-persistence point $(b_v, p_v)$) that was used in the PI's construction. This sum essentially measures how well the node's specific topological feature fits into the overall density landscape of features represented by the PI.
    \item \textbf{Normalization}: The probability measure $\nu_{R_f}(v)$ for node $v$ is obtained by normalizing these contributions across all relevant nodes:
    \[
    \nu_{R_f}(v) := \frac{\text{contrib}(v)}{Z_f}, \quad \text{where the normalization factor is} \quad Z_f = \sum_{u \in V_{R_f}} \text{contrib}(u).
    \]
    This normalization assumes $Z_f > 0$ (i.e., there is at least one node with a non-zero contribution) and ensures that $\sum_{v \in V_{R_f}} \nu_{R_f}(v) = 1$, satisfying the condition for a probability measure.
\end{enumerate}
\end{definition}

This PI-based methodology for assigning node importance offers a more holistic and stable alternative to direct lifespan normalization. Its stability stems from the inherent stability of Persistence Images themselves with respect to perturbations in the input diagrams~\cite{adams2017persistence}. Furthermore, by considering the global distribution of persistence features captured in the PI, it provides a more contextually aware assignment of significance. The formal proof of this measure's stability is a key component of Section~\ref{sec:stability_analysis}.

\section{Stability Analysis of \texorpdfstring{$RGW_p$}{RGWp}}
\label{sec:stability_analysis}

A fundamental requirement for any distance metric intended for robust data analysis is its \textit{stability} with respect to perturbations or noise in the input data. This section is dedicated to formally establishing this crucial property for our proposed Reeb Gromov-Wasserstein distance, $RGW_p(R_f^*, R_g^*)$ (as defined in Definition~\ref{def:rgw_p_method}). Our central aim is to demonstrate that the $RGW_p$ distance between the decorated Reeb graphs $R_f^*$ and $R_g^*$ is controlled by the supremum norm difference, $||f-g||_\infty$, between their respective generating scalar fields, $f$ and $g$.

Our proof strategy, for which full details are provided in Appendix~\ref{app:detailed_stability_proofs}, adopts a hierarchical approach. We establish stability by systematically analyzing the components of the $RGW_p$ distance:
\begin{enumerate}
    \item First, we prove the stability of the metric component used on the Reeb graph nodes, namely the Symmetric Reeb Radius $d_{R_f}$ (Section~\ref{ssec:metric_stability}). 
    \item Next, we establish the stability of the measure component, our novel PI-based probability measure $\nu_{R_f}$ (Section~\ref{ssec:measure_stability}). 
    \item Finally, these individual stability results are integrated to demonstrate the stability of the overall $RGW_p$ distance (Section~\ref{ssec:rgw_stability}).
\end{enumerate}

Throughout this section, we operate under the assumption that the scalar fields $f, g$ are continuous functions defined on topological spaces that are compact, connected, and locally path-connected. For certain results, particularly those concerning the stability of the metric component, we invoke specific regularity conditions known as $(L, \epsilon)$-connectivity (defined below in Definition~\ref{def:stability_L_epsilon_connectivity}), a concept adapted from prior work such as \cite{curry2024stability}.

\subsection{Stability of the Metric Component}
\label{ssec:metric_stability}

The intrinsic distance metric $d_{R_f}$ employed within our $RGW_p$ framework is the Symmetric Reeb Radius, as introduced in Definition~\ref{def:symmetric_reeb_radius}. Establishing its stability draws upon foundational concepts from the metric geometry of fields.

\begin{definition}[M-Metric Field {\cite{curry2024stability}}] \label{def:stability_M_metric_field}
An \textit{M-Metric Field} is a triple $(X, d_X, f)$, where $X$ is a compact, connected, and locally path-connected topological space; $d_X$ is a metric defined on $X$; and $f: X \to M$ is a continuous function mapping $X$ to another metric space $(M, d_M)$. In the context of our work with scalar fields, the target space $M$ is the set of real numbers $\mathbb{R}$, equipped with the standard Euclidean metric $d_M(a,b) = |a-b|$.
\end{definition}

\begin{definition}[Gromov-Hausdorff Distance for M-Metric Fields {\cite{curry2024stability}}] \label{def:stability_GH_M_metric}
Given two M-Metric Fields, $(X, d_X, f)$ and $(Y, d_Y, g)$, their Gromov-Hausdorff Distance, denoted $d_{\mathrm{GH}}((X, f), (Y, g))$, is defined as the infimum of all non-negative values $r \ge 0$ for which an $(r,r)$-correspondence $\EuScript{R} \subset X \times Y$ exists between them. An $(r,s)$-correspondence $\EuScript{R}$ must satisfy two conditions: first, its projections onto $X$ and $Y$ must be surjective; and second, for any two pairs $(x,y) \in \EuScript{R}$ and $(x',y') \in \EuScript{R}$, the distortions are bounded: $|d_X(x,x') - d_Y(y,y')| \le 2r$ and $d_M(f(x),g(y)) \le s$.
\end{definition}

\begin{definition}[$(L, \epsilon)$-Connectivity {\cite{curry2024stability}}] \label{def:stability_L_epsilon_connectivity}
A scalar M-Metric Field $(X, d_X, f)$, where $f:X \to \mathbb{R}$, is said to possess \textit{$(L, \epsilon)$-connectivity} if there exist non-negative constants $L \ge 0$ and $\epsilon \ge 0$ such that for all points $x,y \in X$:
\[ \rho_f(x,y) \le L \cdot d_X(x,y) + 2\epsilon. \]
Here, $\rho_f(x,y) := \inf_{\gamma:x \to y} \sup_{t \in [0,1]} |f(x) - f(\gamma(t))|$ represents the functional variation along paths $\gamma$ within the domain $X$ connecting $x$ to $y$. The class of all such $(L, \epsilon)$-connected fields is denoted by $\EuScript{MF}_{\mathbb{R}}^{L,\epsilon}$.
\end{definition}

\begin{theorem}[Stability of Symmetric Reeb Radius Metric w.r.t. Field GH Distance] \label{thm:stability_drf_gh}
Let $(X,f)$ and $(Y,g)$ be two M-Metric Fields belonging to the class $\EuScript{MF}_{\mathbb{R}}^{L,\epsilon}$. Let $d_{R_f}$ and $d_{R_g}$ be the Symmetric Reeb Radius metrics (Definition~\ref{def:symmetric_reeb_radius}) defined on their respective Reeb graphs $R_f$ and $R_g$. Then, the Gromov-Hausdorff distance between these Reeb graphs (as metric spaces) is bounded as follows:
\[ d_{\mathrm{GH}}((R_f, d_{R_f}), (R_g, d_{R_g})) \leq (L+1) \cdot d_{\mathrm{GH}}((X, f), (Y, g)) + \epsilon. \]
\end{theorem}
\textit{Proof Sketch.}
The proof, detailed in Appendix~\ref{app:proof_thm_stability_drf_gh_content}, adapts a strategy previously employed in \cite{curry2024stability}. The core idea is as follows: Given an $(r,r)$-correspondence $\EuScript{R}$ between the M-Metric Fields $(X,f)$ and $(Y,g)$ (where $r$ is close to $d_{\mathrm{GH}}((X,f),(Y,g))$), we construct a corresponding relation $\EuScript{S}$ between their Reeb graphs $R_f$ and $R_g$. By leveraging the $(L,\epsilon)$-connectivity of the fields and the properties of the correspondence $\EuScript{R}$, we first bound the difference in standard Reeb radii, $|\rho_f(x,x') - \rho_g(y,y')|$, for related pairs of points. This bound on Reeb radii differences is then used to control the distortion of the Symmetric Reeb Radius metrics $d_{R_f}$ and $d_{R_g}$ under the correspondence $\EuScript{S}$. Finally, a standard result from Gromov-Hausdorff theory, which links the distortion of such a correspondence to the Gromov-Hausdorff distance between the metric spaces $(R_f,d_{R_f})$ and $(R_g,d_{R_g})$, yields the stated theorem.
\quad\(\qed\)

\begin{theorem}[Stability of Symmetric Reeb Radius Metric w.r.t. $L_\infty$ Norm] \label{thm:stability_drf_L_infinity}
Let $(X,d_X)$ be a compact, connected, and locally path-connected metric space. Let $f: X \to \mathbb{R}$ and $g: X \to \mathbb{R}$ be two continuous scalar fields such that both $(X,f)$ and $(X,g)$ belong to the class $\EuScript{MF}_{\mathbb{R}}^{L,\epsilon}$. Let $(R_f, d_{R_f})$ and $(R_g, d_{R_g})$ be their respective Reeb graphs equipped with the Symmetric Reeb Radius metric. Then:
\[ d_{\mathrm{GH}}((R_f, d_{R_f}), (R_g, d_{R_g})) \leq (L+1) ||f - g||_\infty + \epsilon, \]
where $||f - g||_\infty = \sup_{x \in X} |f(x) - g(x)|$ is the supremum norm difference between the scalar fields.
\end{theorem}
\textit{Proof Sketch.}
This result is a direct consequence of Theorem~\ref{thm:stability_drf_gh}, specialized to the case where the two scalar fields $f$ and $g$ are defined on the same underlying metric space $(X, d_X)$ (i.e., $Y=X$ and $d_Y=d_X$). The pivotal step is to bound the Gromov-Hausdorff distance between the M-Metric Fields, $d_{\mathrm{GH}}((X,f), (X,g))$, by the $L_\infty$ norm difference $||f-g||_\infty$. This is accomplished by demonstrating that the identity relation $\EuScript{R}_{id} = \{(x,x) \mid x \in X\}$ forms a valid $(r,r)$-correspondence for $r = ||f-g||_\infty$. Substituting this bound for $d_{\mathrm{GH}}((X,f), (X,g))$ into the inequality from Theorem~\ref{thm:stability_drf_gh} directly yields the desired result. The complete argument is provided in Appendix~\ref{app:proof_thm_stability_drf_L_infinity_content}.  
\quad\(\qed\)

\subsection{Stability of the Measure Component}
\label{ssec:measure_stability}

We now turn our attention to the stability of the PI-based probability measure $\nu_{R_f}$ (introduced in Definition~\ref{def:pi_based_measure_method}), which serves as the measure component in our $RGW_p$ framework. The stability of this measure is quantified using the Total Variation distance.

\begin{definition}[Total Variation (TV) Distance] \label{def:stability_tv_distance}
For two probability measures $\mu$ and $\nu$ defined on a common measurable space $(\Omega, \EuScript{F})$, their Total Variation distance, $d_{TV}(\mu, \nu)$, is given by:
\[ d_{TV}(\mu, \nu) = \sup_{A \in \EuScript{F}} |\mu(A) - \nu(A)|. \]
For discrete probability measures defined on a countable space $\Omega_d$, this is equivalent to:
\[ d_{TV}(\mu, \nu) = \frac{1}{2} \sum_{x \in \Omega_d} |\mu(x) - \nu(x)|. \]
\end{definition}

\begin{theorem}[Stability of PI-Based Probability Measure] \label{thm:stability_measure}
Let $f, g: X \to \mathbb{R}$ be continuous scalar fields defined on a compact topological space $X$. Let $\nu_{R_f}$ and $\nu_{R_g}$ be their corresponding PI-based probability measures (Definition~\ref{def:pi_based_measure_method}) on their respective Reeb graphs $R_f$ and $R_g$. Under suitable regularity assumptions on $f$ and $g$ (such as Morse-Smale conditions, which ensure that the finite persistence diagrams and Reeb graph structures behave well under small perturbations) and appropriate choices for the PI construction parameters (e.g., a Lipschitz continuous weighting function and a bounded kernel for the PIs), there exists a constant $M > 0$ such that:
\[ d_{TV}(\nu_{R_f}, \nu_{R_g}) \leq M ||f - g||_\infty. \]
\end{theorem}
\textit{Proof Sketch.}
The proof, for which details can be found in Appendix~\ref{app:proof_thm_stability_measure_content}, proceeds in three main stages: 
\begin{enumerate}
    \item Stability of Extended Persistence Diagrams: It is a well-established result that the bottleneck distance, $d_B$, between the extended persistence diagrams $D(f)$ and $D(g)$ is bounded by the $L_\infty$ norm of the difference between the generating functions: $d_B(D(f),D(g)) \le ||f-g||_\infty$ (as shown in \cite{cohen2009extending}).
    \item Stability of PIs: We then leverage the known stability of Persistence Images with respect to the 1-Wasserstein distance $W_1(D(f),D(g))$ between the input persistence diagrams \cite{adams2017persistence}. Since the $W_1$ distance can itself be bounded by $||f-g||_\infty$ (for instance, $W_1(D(f),D(g)) \le N \cdot ||f-g||_\infty$ for a space decomposable into $N$ cells, as per \cite{skraba2020wasserstein}), it follows that the $L^1$-norm difference $||I_f - I_g||_1$ between the PIs $I_f$ and $I_g$ is also controlled by $||f-g||_\infty$.
    \item Stability of the Normalized Measure $\nu_{R_f}$: Assuming structural stability of the Reeb graphs under small perturbations of the scalar field (which implies a consistent mapping or bijection $\gamma: V_{R_f} \to V_{R_g}$ between the sets of significant nodes), we then bound the difference in contributions, $|\text{contrib}_f(v) - \text{contrib}_g(\gamma(v))|$, using the established stability of the PIs and the Lipschitz properties of the Gaussian kernel used in the PI construction. This subsequently allows us to bound the difference in the normalization factors ($Z_f, Z_g$) and, ultimately, the Total Variation distance $d_{TV}(\nu_{R_f}, \nu_{R_g})$ between the probability measures.
\end{enumerate}
The constant $M$ in the theorem statement depends on various factors, including the parameters chosen for the PI construction (such as kernel bandwidth, weighting function characteristics, and pixel resolution), the complexity of the underlying topological space $X$, and bounds related to the scalar fields $f$ and $g$. 
\quad\(\qed\)

\subsection{Stability of the Reeb Gromov-Wasserstein (\texorpdfstring{$RGW_p$}{RGWp})}
\label{ssec:rgw_stability}

Having established the stability of both the metric component (Symmetric Reeb Radius, Theorem~\ref{thm:stability_drf_L_infinity}) and the measure component (PI-based probability measure, Theorem~\ref{thm:stability_measure}), we are now equipped to combine these results and prove the overall stability of our proposed $RGW_p$ distance (Definition~\ref{def:rgw_p_method}). 

\begin{definition}[Gromov-Prokhorov Distance {\cite{memoli2011gromov}}] \label{def:stability_gp_distance}
Let $X^*=(X, d_X, \mu_X)$ and $Y^*=(Y, d_Y, \mu_Y)$ be two metric measure spaces. Their Gromov-Prokhorov distance, $d_{\EuScript{GP}}(X^*, Y^*)$, is defined as:
\[ \inf \left\{ \varepsilon > 0 \;\middle|\; \exists \pi \in \EuScript{M}(\mu_X, \mu_Y), \exists d \in \EuScript{D}(d_X, d_Y) \text{ s.t. } \pi\{ (x, y) \in X \times Y \mid d(x, y) \ge \varepsilon \} \le \varepsilon \right\}, \]
where $\EuScript{M}(\mu_X, \mu_Y)$ is the set of all couplings between $\mu_X$ and $\mu_Y$ (Definition~\ref{def:coupling_method}), and $\EuScript{D}(d_X, d_Y)$ is the set of all metrics $d$ on the disjoint union $X \sqcup Y$ that extend the original metrics $d_X$ (on $X$) and $d_Y$ (on $Y$).
\end{definition}

\begin{theorem}[Stability of $RGW_p$ Distance] \label{thm:main_rgw_stability}
Let $(X,d_X)$ be a compact, connected, and locally path-connected metric space. Let $f: X \to \mathbb{R}$ and $g: X \to \mathbb{R}$ be continuous scalar fields such that their M-Metric Field representations, $(X,f)$ and $(X,g)$, both belong to the class $\EuScript{MF}_{\mathbb{R}}^{L,\epsilon}$. Let $R_f^*=(R_f, d_{R_f}, \nu_{R_f})$ and $R_g^*=(R_g, d_{R_g}, \nu_{R_g})$ be their respective decorated Reeb graphs, equipped with the Symmetric Reeb Radius metric $d_{R_f}$ (Definition~\ref{def:symmetric_reeb_radius}) and the PI-based probability measure $\nu_{R_f}$ (Definition~\ref{def:pi_based_measure_method}). Furthermore, assume that the diameters of these Reeb graphs, $\text{diam}(R_f, d_{R_f})$ and $\text{diam}(R_g, d_{R_g})$, are uniformly bounded by some constant $D_{max} < \infty$. Then, for any $p \ge 1$, there exist positive constants $C_1$ and $C_2$ (which depend on $L, p, D_{max}$, and parameters arising from the measure stability proof) such that:
\[ RGW_p(R_f^*, R_g^*) \leq C_1 \cdot (||f - g||_\infty)^{1/p} + C_2 \cdot \epsilon^{1/p}. \]
In particular, for the case $p=1$:
\[ RGW_1(R_f^*, R_g^*) \leq C_1' \cdot ||f - g||_\infty + C_2' \cdot \epsilon, \]
for appropriately defined constants $C_1'$ and $C_2'$.
\end{theorem}
\textit{Proof Sketch.}
The proof, provided in detail in Appendix~\ref{app:proof_thm_main_rgw_stability_content}, hinges on a key inequality established by Mémoli~\cite{memoli2011gromov} that relates the $RGW_p$ distance to the Gromov-Prokhorov distance $d_{\EuScript{GP}}$: specifically, $RGW_p(R_f^*, R_g^*) \le (d_{\EuScript{GP}}(R_f^*, R_g^*))^{1/p} (D_{max}^p+1)^{1/p}$. The main steps are as follows:
\begin{enumerate}
    \item First, we bound the Gromov-Hausdorff distance between the metric structures of the Reeb graphs, $d_{\mathrm{GH}}((R_f,d_{R_f}),(R_g,d_{R_g}))$, by $(L+1)||f-g||_\infty + \epsilon$, using our earlier result from Theorem~\ref{thm:stability_drf_L_infinity}. Let this bound be denoted by $\delta_{GH}$.
    \item Second, we bound the Total Variation distance between the probability measures on the Reeb graphs, $d_{TV}(\nu_{R_f},\nu_{R_g})$, by $M||f-g||_\infty$, using Theorem~\ref{thm:stability_measure}. Let this bound be $\eta_{TV}$.
    \item With these bounds on the geometric and measure discrepancies, we then establish an upper bound for the Gromov-Prokhorov distance $d_{\EuScript{GP}}(R_f^*,R_g^*)$. Standard results in the theory of mm-spaces indicate that $d_{\EuScript{GP}}(R_f^*,R_g^*) \le \max(\delta_{GH}, \eta_{TV})$.
    \item Finally, this upper bound for $d_{\EuScript{GP}}$ is substituted into Mémoli's inequality. Subsequent algebraic manipulation, utilizing properties such as $\max(A+B,C) \le \max(A,C)+B$ and $(a+b)^{1/p} \le a^{1/p} + b^{1/p}$ (for non-negative $a,b,A,B,C$ and $p \ge 1$), leads to the final inequality form. The constants $C_1$ and $C_2$ consolidate the various terms involving $L, M, D_{max}$, and $p$.
\end{enumerate}
This chain of reasoning demonstrates that small perturbations in the input scalar field (as measured by $||f-g||_\infty$) and a small $\epsilon$ value (related to the $(L,\epsilon)$-connectivity assumption) indeed lead to correspondingly small changes in the computed $RGW_p$ distance, thereby establishing the desired stability.
\quad\(\qed\)

\section{An Illustrative Example: Computing \texorpdfstring{$RGW_p$}{RGWp} Between Two Reeb Graphs}
\label{ssec:example_rgw_computation}

\begin{figure}[bt!]
\centering
\includegraphics[scale=0.85]{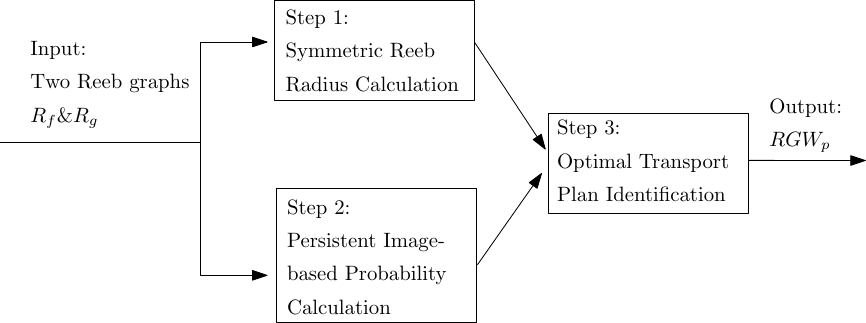} 
\caption{The computational workflow for calculating the $RGW_p$ distance between two Reeb graphs. This process involves three key steps: Symmetric Reeb Radius calculation for intra-graph distances, PI-based probability measure determination for node weighting, and finally, the Optimal Transport computation to find the GW distance.}
\label{fig:workflow}
\end{figure}

This section serves to illustrate the practical computation of our proposed Reeb Gromov-Wasserstein distance, $RGW_p$, by walking through an example comparing two specific Reeb graphs, denoted $R_f$ and $R_g$. As conceptually depicted in the workflow diagram (Figure~\ref{fig:workflow}), the overall process unfolds in three primary stages:
\begin{itemize}
    \item[\textbf{(i)}] Computation of the intra-graph distance matrices, $d_{R_f}$ and $d_{R_g}$, using the Symmetric Reeb Radius.
    \item[\textbf{(ii)}] Determination of the PI-based Borel probability measures, $\nu_{R_f}$ and $\nu_{R_g}$, assigned to the nodes of each respective graph.
    \item[\textbf{(iii)}] Identification of the optimal transport (OT) plan, $\pi^*$, between these derived measures and the subsequent computation of the final $RGW_p$ distance value.
\end{itemize}

\begin{figure*}[tb!]
    \centering
    \begin{subfigure}[t]{0.49\textwidth}
        \centering
        \includegraphics[width=\textwidth]{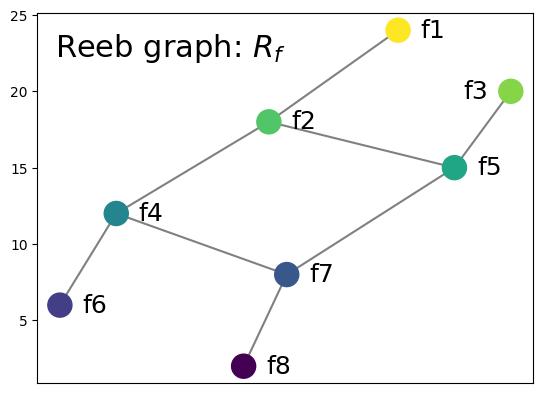} 
    \end{subfigure}
    \hfill 
    \begin{subfigure}[t]{0.49\textwidth}
        \centering
        \includegraphics[width=\textwidth]{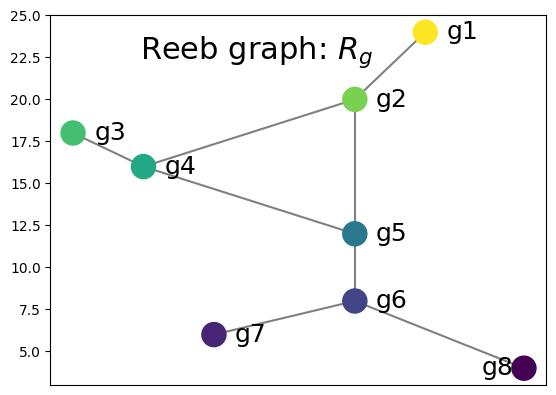} 
    \end{subfigure}
    \caption{The two example Reeb graphs utilized for the $RGW_p$ computation: $R_f$ (left) comprises nodes labeled $f_1, \dots, f_8$, and $R_g$ (right) consists of nodes $g_1, \dots, g_8$. These node labels are used consistently in the subsequent presentation of distance matrices and probability vectors.}
    \label{fig:RGW_p_example_graphs}
\end{figure*}

\textbf{(i) Step 1: Symmetric Reeb Radius Calculation.}
We commence with the two example Reeb graphs, $R_f$ and $R_g$, which are illustrated in Figure~\ref{fig:RGW_p_example_graphs}. For each graph, we compute the Symmetric Reeb Radius $d_{R_f}(v,u)$ (as defined in Definition~\ref{def:symmetric_reeb_radius}) for all pairs of nodes $(v,u)$ within that graph. This calculation first involves determining the standard (asymmetric) Reeb radii $\rho_f(v,u)$ and $\rho_f(u,v)$, which are then symmetrized using the formula $d_{R_f}(v,u) = \frac{1}{2}(\rho_f(v,u) + \rho_f(u,v))$. For our chosen example graphs $R_f$ and $R_g$, this process yields the following symmetric distance matrices (where only the upper triangles are explicitly shown, given their symmetry):
\begin{equation*} 
\setlength{\arraycolsep}{2pt} 
\begin{array}{cc}
d_{R_f} =

\scalebox{0.8}{$ 

\begin{pmatrix}

0 & 6 & 7 & 12 & 9 & 18 & 16 & 22 \\

 & 0 & 4 & 6 & 3 & 12 & 10 & 16 \\

 & & 0 & 8 & 5 & 14 & 12 & 18 \\

 & & & 0 & 3.5 & 6 & 4 & 10 \\

& & & & 0 & 9 & 7 & 13 \\

 & & & & & 0 & 5 & 8 \\

 & & & & & & 0 & 6 \\

\multicolumn{8}{c}{\text{Symm.}} 

\end{pmatrix}

$}

, &

d_{R_g} =

\scalebox{0.8}{$

\begin{pmatrix}

0 & 4 & 7 & 8 & 12 & 16 & 18 & 20 \\

 & 0 & 3 & 4 & 8 & 12 & 14 & 16 \\

 & & 0 & 2 & 6 & 10 & 12 & 14 \\

& & & 0 & 4 & 8 & 10 & 12 \\

 & & & & 0 & 4 & 6 & 8 \\

& & & & & 0 & 2 & 4 \\

& & & & & & 0 & 3 \\

\multicolumn{8}{c}{\text{Symm.}}
\end{pmatrix}
$}
\end{array}
\end{equation*}
The entries in the lower triangle are omitted due to symmetry (i.e., $d(u,v) = d(v,u)$), and all diagonal entries are zero.

\begin{figure}[bt!]
\centering
\includegraphics[scale=0.34]{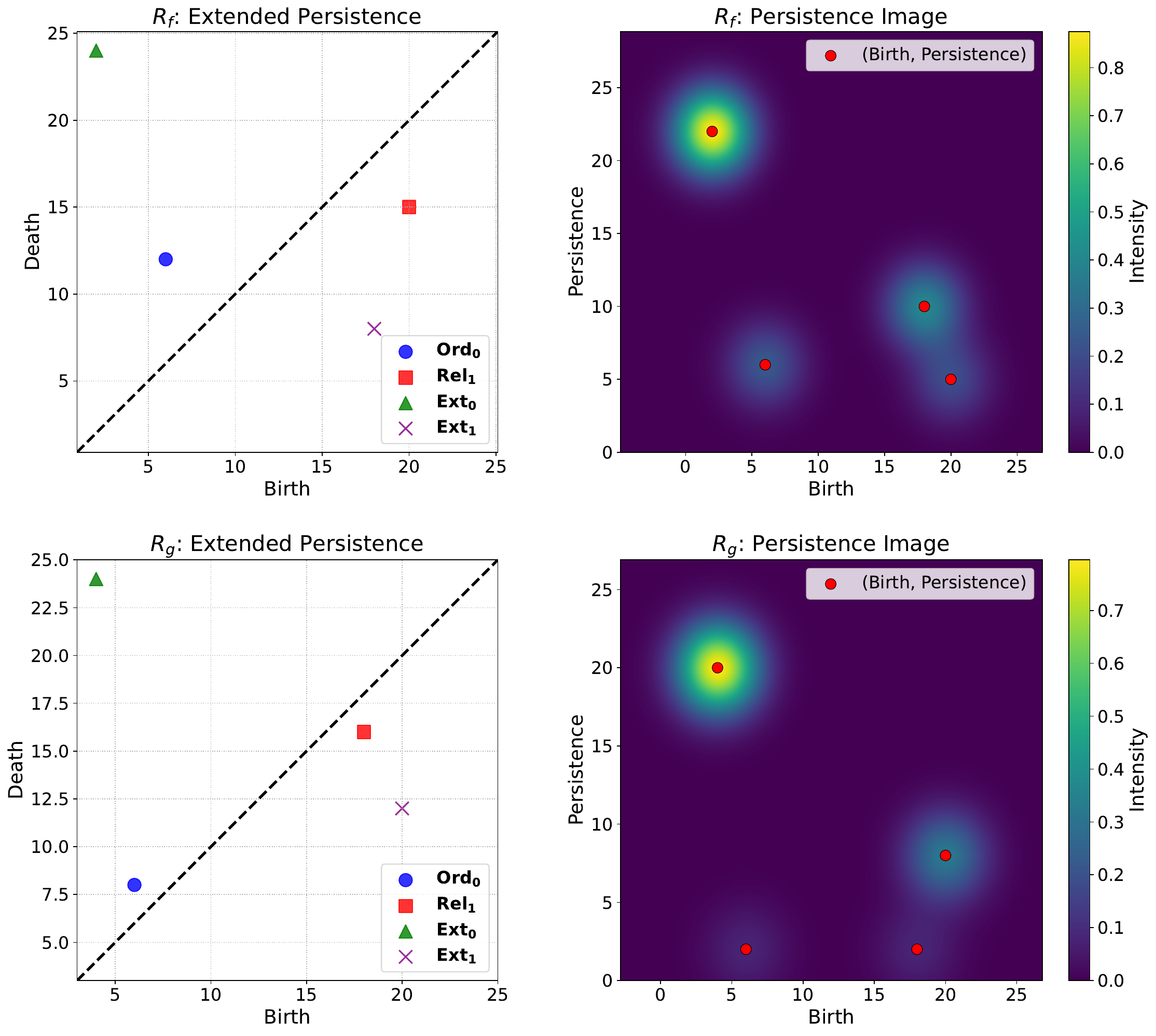} 
\caption{A conceptual illustration depicting the extended persistence diagrams (PDs, shown in the left column) and their corresponding Persistence Images (PIs, right column) for the example Reeb graphs $R_f$ (top row) and $R_g$ (bottom row). Points within the PDs are color-coded by their type ($Ord_0, Rel_1, Ext_0, Ext_1$). The PIs, which represent feature density in the birth-persistence plane, form the basis for deriving the node probability measures.}
\label{fig:example_pd_pi}
\end{figure}

\textbf{(ii) Step 2: PI-Based Probability Measure Calculation.}
Next, we determine the Borel probability measures, $\nu_{R_f}$ on $V_{R_f}$ and $\nu_{R_g}$ on $V_{R_g}$, utilizing our PI-based approach (Definition~\ref{def:pi_based_measure_method}). For each Reeb graph, its extended persistence diagram ($D_f$ and $D_g$, respectively) is computed by standard software packages like GUDHI \cite{maria2014gudhi}. This step identifies the key topological features (classified as $Ord_0, Rel_1, Ext_0, Ext_1$) and, crucially, their associated Reeb graph nodes.
For our example graph $R_f$, the diagram $D_f$ reveals an $\mathbf{Ord_0}$ feature corresponding to the node pair $(f_6,f_4)$, a $\mathbf{Rel_1}$ feature associated with nodes $(f_3,f_5)$, the primary $\mathbf{Ext_0}$ feature pairing nodes $(f_8,f_1)$, and an $\mathbf{Ext_1}$ feature linked to nodes $(f_2,f_7)$.
Similarly, for graph $R_g$, its diagram $D_g$ identifies an $\mathbf{Ord_0}$ feature pairing $(g_7,g_6)$, a $\mathbf{Rel_1}$ feature with $(g_3,g_4)$, the main $\mathbf{Ext_0}$ feature pairing $(g_8,g_1)$, and an $\mathbf{Ext_1}$ feature associated with nodes $(g_2,g_5)$.
These persistence diagrams are then transformed into their respective Persistence Images ($I_f, I_g$), a process conceptually illustrated in Figure~\ref{fig:example_pd_pi}. The node contributions are calculated from these PIs as per Definition~\ref{def:pi_based_measure_method} and subsequently normalized to yield the probability measures.

For the Reeb graph $R_f$, the resulting probability measure $\nu_{R_f}$ assigns the following approximate probabilities to its nodes $(f_1, \dots, f_8)$ respectively:
\[
\nu_{R_f} \approx (0.2420, 0.1190, 0.0730, 0.0660, 0.0730, 0.0660, 0.1190, 0.2420).
\]
For the Reeb graph $R_g$, the node probabilities for its nodes $(g_1, \dots, g_8)$ are similarly found to be:
\[
\nu_{R_g} \approx (0.3075, 0.1255, 0.0385, 0.0385, 0.1255, 0.0284, 0.0284, 0.3075).
\]
It is noteworthy that in both cases, nodes such as $f_1, f_8$ (for $R_f$) and $g_1, g_8$ (for $R_g$), which are associated with highly persistent $Ext_0$ features (often global minimum/maximum), typically receive higher probability weights, reflecting their significant topological role.

\textbf{(iii) Step 3: Optimal Transport Plan and $RGW_p$ Computation.}
Finally, armed with the intra-graph distance matrices $d_{R_f}$ and $d_{R_g}$, and the node probability measures $\nu_{R_f}$ and $\nu_{R_g}$, we proceed to compute the $RGW_p$ distance itself. This requires finding an OT plan, denoted $\pi^* \in \EuScript{M}(\nu_{R_f}, \nu_{R_g})$, that minimizes the cost function specified in our Definition~\ref{def:rgw_p_method}. For this particular illustrative example, we set $p=2$.  While the exact computation of the GW distance can be NP-hard, its practical calculation, as performed in this work, leverages efficient iterative algorithms designed to find high-quality approximate solutions. Such computations can be efficiently handled using existing libraries like the Python Optimal Transport ~\cite{flamary2021pot}, which takes $d_{R_f}$, $d_{R_g}$, $\nu_{R_f}$, and $\nu_{R_g}$ as inputs. The solver then outputs both the (approximate) optimal transport plan $\pi^*$ and the resulting $RGW_p$ distance.

\begin{figure}[bt!]
\centering
\includegraphics[scale=0.32]{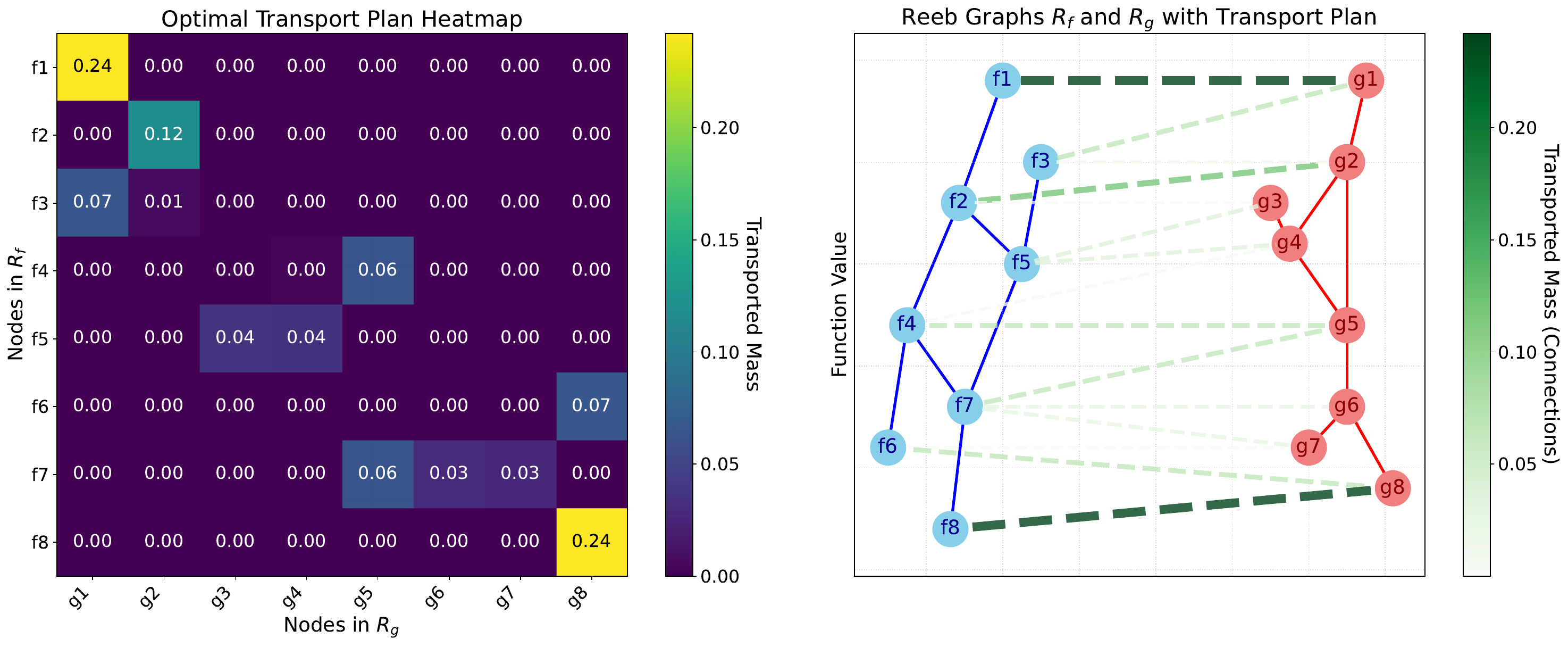} 
\caption{(a) A heatmap representation of the optimal transport plan $\pi^*$ computed for $RGW_2(R_f^*, R_g^*)$. The intensity of each cell $(v,w)$ corresponds to $\pi^*_{vw}$, representing the amount of mass transported between node $v \in V_{R_f}$ and node $w \in V_{R_g}$. (b) A visualization highlighting the prominent node-to-node correspondences between $R_f$ and $R_g$ as implied by the optimal plan $\pi^*$.}
\label{fig:heatmap_ot_plan_example}
\end{figure}

The computed optimal transport plan $\pi^*$ for this example is visualized as a heatmap in Figure~\ref{fig:heatmap_ot_plan_example}(a), with Figure~\ref{fig:heatmap_ot_plan_example}(b) offering a more direct visualization of the key node-to-node correspondences implied by this plan. This optimal plan $\pi^*$ inherently satisfies the marginal constraints imposed by $\nu_{R_f}$ and $\nu_{R_g}$. The final computed Reeb Gromov-Wasserstein distance for this example is $RGW_2(R_f^*, R_g^*) = 7.814688$. This value quantitatively expresses the dissimilarity between the metric measure structures of $R_f^*$ and $R_g^*$.

\section{Experiments}
\label{sec:experiments} 

In this section, we present a series of empirical evaluations designed to assess the performance of our proposed Reeb Gromov-Wasserstein distance, $RGW_p$. The primary objective is to compare $RGW_p$ against established metrics in the context of a 3D point cloud classification task. Furthermore, we conduct ablation studies to investigate the contributions of the different components within the $RGW_p$ framework and analyze the sensitivity of our method to key hyperparameters in the Persistence Image construction. Our source code is publicly available at our GitHub repository.\footnote{https://github.com/gm3g11/Gromov-Wasserstein-distance-on-Reeb-graphs}

\subsection{Experimental Setup}
\label{ssec:eval_metric_setup}

To ensure a rigorous evaluation, we detail our experimental design, including the datasets, Reeb graph construction pipeline, benchmark algorithms, evaluation metrics, and the hardware platform used.

\begin{table}[hbt!]
\centering
\caption{Point cloud datasets used in the experiments. Note: Modelnet10$^\dagger$ includes only 25\% of the original data due to memory constraints.\label{tab:dataset}}
\begin{tabular*}{\textwidth}{l @{\extracolsep{\fill}} ccc l} 
\toprule
Dataset & Size & Queries & Labels & Description \\ 
\midrule
Modelnet10$^\dagger$ & 1000 & 200 & 10 & CAD models of 10 household object categories. \\ 
SHREC14 & 320 & 80 & 40 & Non-rigid 3D human models from retrieval challenge. \\ 
Mesh & 100 & 10 & 7 & Various triangulated 3D mesh objects. \\
\bottomrule
\end{tabular*}
\end{table}

\textbf{Datasets.} We evaluate the proposed metrics on three widely used datasets (summarized in Table \ref{tab:dataset}): ModelNet10 \cite{wu20153d}, a curated collection of CAD models representing household objects; SHREC14 \cite{godil2014shrec}, a dataset designed for 3D shape retrieval challenges; and the Triangulated Mesh Database \cite{sumnermesh}, which contains triangulated 3D meshes of various objects. Note that we randomly selected only $25\%$ data of the ModelNet10 dataset due to the DRGs metric encountering out-of-memory issues; the other compared metrics do not experience this memory limitation.

\begin{figure}[hbt]
\centering
\includegraphics[ width=0.9\textwidth]{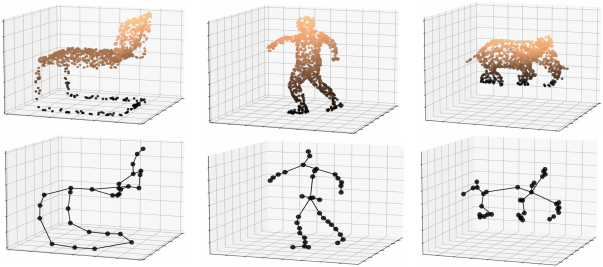}

\caption{Examples of 3D point clouds (top row) and their corresponding mapper graphs (bottom row) from datasets including ModelNet10, SHREC14, and Mesh.}
\label{fig:generate}
\end{figure}

\textbf{Mapper Graph Construction  (Reeb Graph Approximation).}
To compute the proposed \( RGW_p \), constructing a representation of the Reeb graph is a necessary preliminary step. However, computing the exact theoretical Reeb graph can be computationally intensive for complex data, leading to the common use of approximations like Mapper graphs \cite{alvarado2024any, singh2007topological}. The construction process begins by uniformly sampling 1024 points from each dataset to create a 3D point cloud representation. Subsequently, the \textit{p-eccentricity filtration function} is employed to generate a Vietoris-Rips complex \cite{memoli2011gromov}. Finally, the Mapper algorithm \cite{singh2007topological} is applied to the Vietoris-Rips complex to produce a Mapper graph, which serves as a discrete approximation of the theoretical Reeb graph. Figure~\ref{fig:generate} illustrates examples of the 3D point clouds and their corresponding Mapper graphs.

\textbf{Algorithms and Implementations.} In the proposed $RGW_p$ metric, we utilize the GUDHI package \cite{maria2014gudhi} to compute the extended persistence diagrams and the POT library \cite{flamary2021pot} to calculate the GW distance. For benchmarking, we compare our method against the bottleneck distance, graph edit distance (GED), and DRGs distance with persistent homology decoration \cite{curry2023topologically}. The bottleneck distance is computed using the Ripser package \cite{ctralie2018ripser}. Since GED is known to be an NP-hard problem \cite{di2016edit}, we adopt an approximate GED implementation~\cite{abu2015exact}, leveraging the NetworkX package \cite{hagberg2008exploring} for practical computations.

\textbf{Evaluation Metric.}
The primary task for evaluating the performance of these distance metrics is 3D point cloud classification using a k-Nearest Neighbors (k-NN) approach. We adopt \textbf{recall@k accuracy} as our main evaluation metric. The recall@k for a given query shape is defined as the fraction of queries for which at least one of its true class nearest neighbors (from the same category) is found within the top-$k$ closest shapes retrieved by the distance metric. The overall recall@k accuracy is the average over all queries.
In addition to classification accuracy, we also report the \textbf{execution time} for each method, defined as the total time required to compute all pairwise distances for the queries within each dataset.

\textbf{Hardware platform.}
We run the implementations on an Intel i9-12900K CPU with 128 GB of DDR4 memory.

\subsection{Comparison of Metrics on 3D Point Cloud } \label{ssec:main_comparison_results}

\begin{figure}[bt]
\centering
\includegraphics[ width=1.0\textwidth]{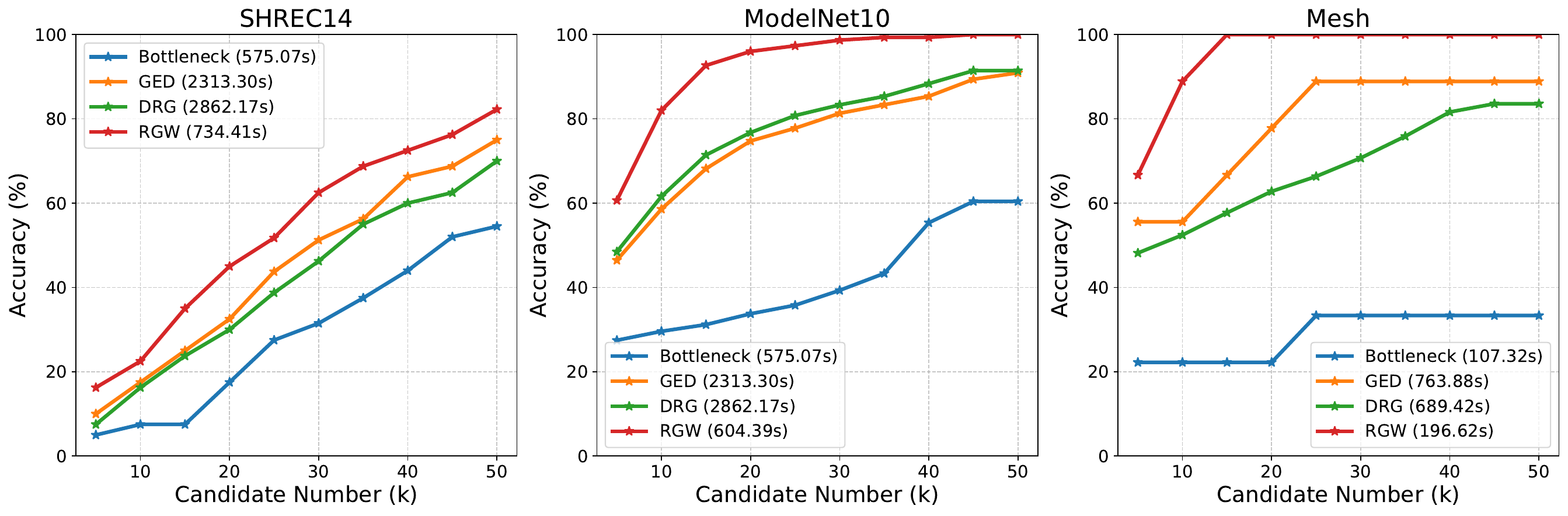}

\captionsetup{format=plain, width=1.0\textwidth, font=small, labelfont=bf}
\caption{Comparison of k-NN classification accuracy (\%) across different methods (Bottleneck, GED, DRG, RGW) on SHREC14, ModelNet10, and Mesh datasets. Each subplot plots accuracy against the number of candidates (k), with method labels indicating execution time in seconds.}

\label{fig:metric_comparison}

\end{figure}

Figure~\ref{fig:metric_comparison} presents a comparative evaluation of four methods---Bottleneck distance, GED (Graph Edit Distance), DRG (Decorated Reeb Graphs), and the proposed $RGW_p$ (Reeb Gromov-Wasserstein)---across three datasets: SHREC14, ModelNet10, and Mesh. The evaluation focuses on k-NN classification accuracy, with execution times for each method also noted in the legend (e.g., RGW takes $734.41$s on SHREC14, $604.39$s on ModelNet10, and $196.62$s on Mesh). Each of the three subplots corresponds to one dataset and plots the accuracy against the number of candidates ($k$) ranging from 5 to 50.

Across all datasets, the proposed $RGW_p$ method consistently demonstrates superior performance. It achieves the highest final accuracies at $k=50$: $82.25\%$ on SHREC14, $100\%$ on ModelNet10, and $100\%$ on Mesh. On ModelNet10 and Mesh, $RGW_p$ not only reaches perfect accuracy but generally does so more rapidly as $k$ increases. For instance, on ModelNet10, $RGW_p$ surpasses $90\%$ accuracy by $k=15$ (achieving $92.67\%$) and reaches $100\%$ by $k=45$. This is a level only approached by GED (reaching $90.91\%$ at $k=50$) and DRG (reaching $91.43\%$ at $k=50$) at higher $k$ values. Similarly, on the Mesh dataset, $RGW_p$ achieves $100\%$ accuracy by $k=15$, significantly outperforming GED, which plateaus around $88.89\%$, and DRG, which reaches approximately $83.58\%$ at $k=50$.

For the more challenging SHREC14 dataset, $RGW_p$ also shows a clear advantage, maintaining the highest accuracy curve throughout the range of $k$ and reaching $82.25\%$ at $k=50$. This compares favorably to GED's $75.00\%$, DRG's $70.00\%$, and the Bottleneck distance's $54.50\%$ at the same $k$ value. While GED and DRG can achieve competitive accuracies, especially on ModelNet10, their reported execution times (e.g., GED: $2313.3$s, DRG: $2862.17$s for the respective datasets shown in the legend) are substantially higher than those for $RGW_p$. The Bottleneck distance, while consistently the fastest method, yields the lowest classification accuracies across all datasets and $k$ values. These results highlight that $RGW_p$ offers a strong balance, achieving high classification accuracy while maintaining competitive computational efficiency compared to other structural graph comparison methods.

\subsection{Evaluation of Proposed Components in \texorpdfstring{$RGW_p$}{RGWp}}
\begin{table*}[hbt]
\centering
\caption{Ablation study results for the $RGW_p$ framework on the ModelNet10, SHREC14, and Mesh datasets. The table shows k-NN (for $k=20$) classification accuracies for three experimental groups: Group 1 varies the Reeb graph distance metric (replacing our proposed Symmetric Reeb Radius while keeping the PI-based probability measure); Group 2 varies the probability measure (replacing our PI-based measure while keeping the Symmetric Reeb Radius); Group 3 (Proposed) uses both our proposed Symmetric Reeb Radius and PI-based probability measure.}
\label{tab:ablation_study_components} 
\begin{tabular}{l|ccc} 
\hline
\textbf{Distance/Weight Type} & \textbf{Modelnet10} & \textbf{SHREC14} & \textbf{Mesh} \\ \hline
Reeb Radius & 92.64\% & 38.35\% & 94.83\% \\  
Reeb Distance & 88.27\% & 36.29\% & 90.89\% \\  
Shortest Path & 91.86\% & 39.1\% & 93.62\% \\  
Max-based Symmetric Reeb Radius \cite{curry2024stability} & 93.41 \% & 40.23\%  & 95.47\% \\ \hline
Uniform & 70.42\% & 17.83\%  & 72.97\% \\  
Intensity-based &68.83\% & 15.91\%   & 70.17\% \\  
Degree-based & 67.97\% & 16.93\%   &73.41\% \\  
Lifespan-based & 76.75\% & 23.5\%   &79.41\% \\  
\hline
Proposed & \textbf{96.16\%} & \textbf{45.38\%} & \textbf{100\%}\\  
\hline
\end{tabular}
\end{table*}

Table~\ref{tab:ablation_study_components} presents an ablation study to evaluate the impact of the core components of our proposed $RGW_p$ framework: the Symmetric Reeb Radius and the PI-based probability measure. The experiments are conducted on the ModelNet10, SHREC14, and Mesh datasets, reporting k-NN classification accuracy.

The first group of experiments (Group 1) assesses alternative distance metrics for Reeb graph nodes while keeping our PI-based probability measure fixed. The metrics evaluated include the standard (asymmetric) Reeb Radius, Reeb Distance, Shortest Path distance, and the Max-based Symmetric Reeb Radius from \cite{curry2024stability}. Among these, the Max-based Symmetric Reeb Radius generally performs best, achieving accuracies such as $93.41\%$ on ModelNet10 and $95.47\%$ on Mesh. However, our proposed $RGW_p$ (Group 3), which uses the (average-based) Symmetric Reeb Radius, consistently outperforms all variants in this group across all datasets, achieving $96.16\%$ on ModelNet10, $45.38\%$ on SHREC14, and $100\%$ on Mesh. This suggests that while symmetrizing the Reeb radius is beneficial, our averaging approach is more effective than the maximum-based alternative within our framework.

The second group of experiments (Group 2) evaluates different probability measures while using our proposed Symmetric Reeb Radius as the fixed distance metric. The alternatives include a Uniform distribution, an Intensity-based measure, a Degree-based measure, and a Lifespan-based measure derived directly from persistence values. The results show that these alternative measures generally lead to significantly lower accuracies compared to our PI-based approach. For instance, on ModelNet10, the Uniform ($70.42\%$), Intensity-based ($68.83\%$), Degree-based ($67.97\%$), and even the Lifespan-based ($76.75\%$) measures are substantially outperformed by our proposed method's $96.16\%$. A similar trend is observed on SHREC14 (e.g., Lifespan-based at $23.5\%$ vs. Proposed at $45.38\%$) and Mesh (Lifespan-based at $79.41\%$ vs. Proposed at $100\%$). This underscores the significant contribution of the PI-based probability measure, which better captures topological importance and leads to more discriminative Reeb graph comparisons.

Overall, the results in Table~\ref{tab:ablation_study_components} (Group 3) confirm that the combination of our proposed Symmetric Reeb Radius and the PI-based probability measure yields the best performance, validating the design choices for our $RGW_p$ framework.

\subsection{Hyperparameter Analysis for Persistence Image Construction}
\label{ssec:hyperparameter_analysis_pi}

\begin{figure}[bt]
\centering
\includegraphics[ width=1.0\textwidth]{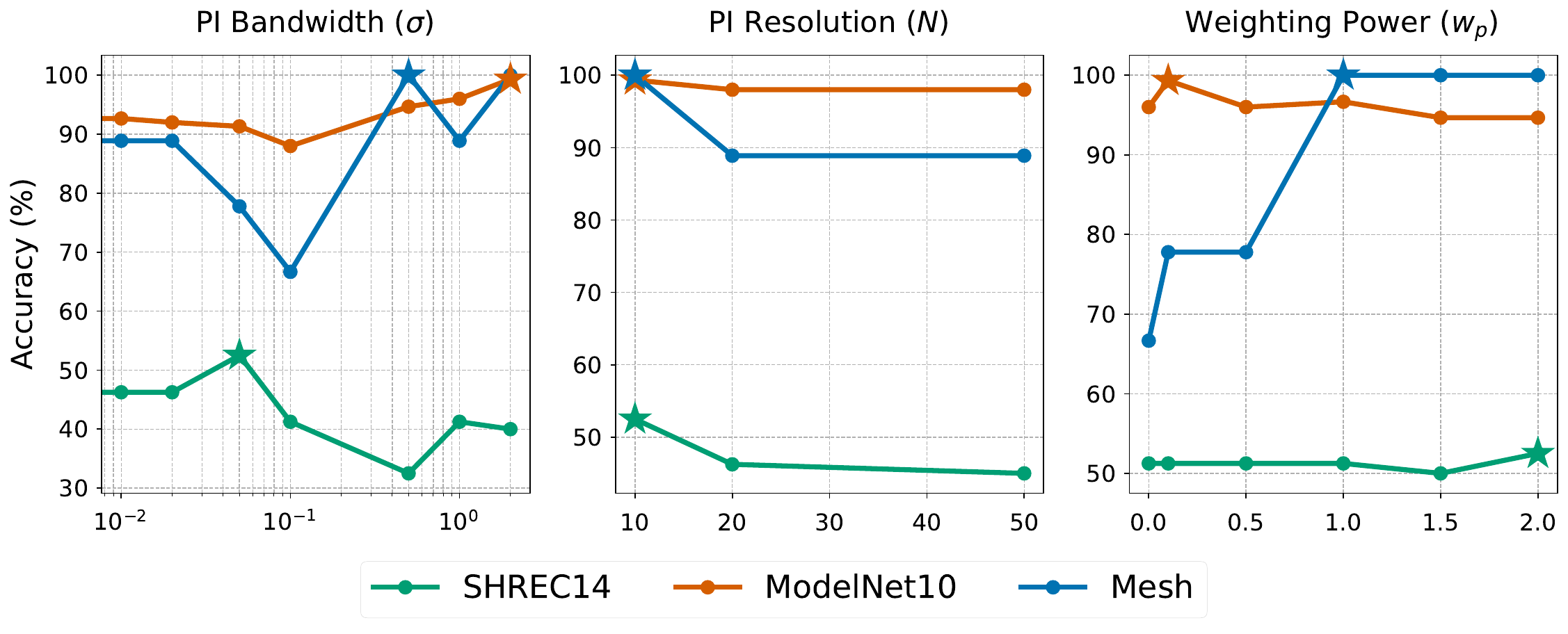}

\captionsetup{format=plain, width=1.0\textwidth, font=small, labelfont=bf}
\caption{Sensitivity analysis of the core Persistence Image hyperparameters. Each plot illustrates the classification accuracy (at k=20) on the SHREC14, ModelNet10, and Mesh datasets as a single parameter is varied: (left) PI Bandwidth ($\sigma$), (middle) PI Resolution ($N$), and (right) Weighting Power ($w_p$). For each curve, the other parameters are held constant at their optimal values. The optimal setting for the varied parameter is marked with a star.}

\label{fig:sensitivity_anlysis}

\end{figure}

In our proposed $RGW_p$ method, the primary hyperparameters are inherited from the Persistence Image (PI) construction, as detailed in Section~\ref{sec:persistence_images_bg}. These three key parameters are: the Gaussian kernel bandwidth ($\sigma$), the PI grid resolution ($N$), and the weighting power ($w_p$) applied to persistence values. To ensure a robust evaluation of our method, this section presents a thorough sensitivity analysis of these three hyperparameters across our test datasets.

Our analysis began with a comprehensive grid search to explore the parameter space. The ranges for each parameter were: PI Bandwidth ($\sigma$) in $\{0, 0.01, 0.02, 0.05, 0.1, 0.5, 1, 2\}$; PI Resolution ($N$) in $\{10, 20, 50\}$; and Weighting Power ($w_p$) in $\{0, 0.1, 0.5, 1, 1.5, 2\}$.

To determine the optimal configuration for each dataset, we used the k-NN classification accuracy at $k=20$ as our primary evaluation metric. This value was chosen as it provided better differentiation between high-performing models compared to metrics at higher $k$ values, which tended to be saturated. In cases where multiple parameter sets achieved the same top accuracy, we selected the most computationally efficient configuration (i.e., the one with the lowest PI Resolution $N$). The resulting optimal parameters are ($\sigma=0.05, N=10, w_p=2$) for SHREC14, ($\sigma=2, N=10, w_p=0.1$) for ModelNet10, and ($\sigma=0.5, N=10, w_p=1$) for Mesh.

To visualize the stability of our method around these optimal points, we plotted the accuracy while varying a single parameter and holding the other two constant at their best-found values. The results are presented in Figure~\ref{fig:sensitivity_anlysis}. Our analysis of the figure reveals several key insights into the behavior of the hyperparameters.

\textbf{PI Bandwidth ($\sigma$):} This parameter demonstrates the most varied influence. For SHREC14, a small, non-zero bandwidth of $\sigma=0.05$ is clearly optimal, suggesting that a slight blurring of features is beneficial while too much is detrimental. For the Mesh dataset, performance peaks at a moderate $\sigma=0.5$ but degrades significantly at lower values before recovering. In contrast, ModelNet10 achieves its best results at a high $\sigma=2$, indicating a high degree of tolerance to feature smoothing. This suggests that the optimal level of PI blurring is highly dependent on the geometric and topological characteristics of the underlying dataset.

\textbf{PI Resolution ($N$):} The effect of resolution is remarkably consistent across all three datasets. Optimal or near-optimal performance is achieved at the lowest resolution tested ($N=10$). As resolution increases, accuracy tends to either plateau or slightly decrease. This is a significant finding, as it implies that a coarse grid is not only sufficient but often preferable, potentially by avoiding the overfitting that can occur with finer grids. This result also has positive practical implications, as using a lower resolution significantly reduces the computational cost of both generating and comparing Persistence Images.

\textbf{Weighting Power ($w_p$):} The influence of the weighting power, which controls the emphasis on high-persistence features, also varies. SHREC14 benefits from a high power ($w_p=2$), indicating its classification relies heavily on the most prominent topological features. Conversely, ModelNet10 performs best with a very low power ($w_p=0.1$), suggesting that all features, including those with low persistence, contribute valuable information. The Mesh dataset finds a balance at $w_p=1$. This variation implies that the ideal approach to weighting topological features is dataset-specific, depending on the relative importance of major structural components versus finer-grained details.

\section{Conclusion}
\label{sec:conclusion}

This paper introduced $RGW_p$, a novel Gromov-Wasserstein distance specifically engineered for robust Reeb graph comparison. Our framework's distinctiveness stems from its synergistic integration of a Symmetric Reeb Radius for stable intra-graph distance calculations and a novel PI-based probability measure that effectively imbues Reeb graph nodes with topologically significant weights. This design allows $RGW_p$ to faithfully capture the complex geometric and topological features of scalar fields.

A cornerstone of our work is the rigorous theoretical proof establishing the $L_\infty$-stability of $RGW_p$ against perturbations in the underlying scalar field, a critical guarantee for reliable analysis in practical settings. Extensive experimental evaluations further confirmed the practical advantages of $RGW_p$, demonstrating its superior performance in classification accuracy and competitive computational efficiency when compared against traditional Reeb graph comparison techniques. Consequently, this research delivers $RGW_p$ as a potent, provably stable, and empirically validated new tool for advancing topological data analysis and the comparative study of scalar fields.


\bibliography{main}

\newpage
\appendix

\section{Proof of Proposition~\ref{prop:metric}}
\label{appendix:proof}

\begin{proof}
Let $x, y, z$ be arbitrary nodes in the node set $V_{R_f}$ of a Reeb graph $R_f$. We aim to demonstrate that the Symmetric Reeb Radius, $d_{R_f}(x,y) = \frac{1}{2}(\rho_f(x,y) + \rho_f(y,x))$, where $\rho_f$ is the Reeb Radius (Definition~\ref{def:reeb_radius}), satisfies the properties of a metric.

\begin{enumerate}
    \item \textbf{Non-negativity:}
    The Reeb radius $\rho_f(x, y)$ is defined as an infimum of suprema of absolute differences of function values, specifically $\sup_{t \in [0, 1]} |\bar{f}(x) - \bar{f}(\gamma(t))|$ for paths $\gamma$. Since absolute values are inherently non-negative, their supremum over any path is also non-negative. Consequently, the infimum of these non-negative values, $\rho_f(x, y)$, must satisfy $\rho_f(x, y) \geq 0$. Therefore, $d_{R_f}(x, y) = \frac{1}{2}(\rho_f(x, y) + \rho_f(y, x))$ must also be non-negative.

    \item \textbf{Identity of Indiscernibles:}
    If $x = y$, then the path $\gamma$ from $x$ to $x$ can be the trivial path consisting only of the node $x$. For this path, $\bar{f}(\gamma(t)) = \bar{f}(x)$ for all $t$, so $\sup_{t \in [0, 1]} |\bar{f}(x) - \bar{f}(\gamma(t))| = 0$. Thus, $\rho_f(x, x) = 0$. This implies $d_{R_f}(x, x) = \frac{1}{2}(\rho_f(x, x) + \rho_f(x, x)) = 0$.

    Conversely, assume $d_{R_f}(x, y) = 0$. Given that $\rho_f(x,y)$ and $\rho_f(y,x)$ are non-negative, this equality implies that both $\rho_f(x, y) = 0$ and $\rho_f(y, x) = 0$.
    The condition:
    
    $\rho_f(x, y) = \inf_{\gamma: x \to y} \sup_{t \in [0,1]} |\bar{f}(x) - \bar{f}(\gamma(t))| = 0$ means that for any $\delta > 0$, there exists a path $\gamma$ from $x$ to $y$ such that $\sup_{t \in [0,1]} |\bar{f}(x) - \bar{f}(\gamma(t))| < \delta$. This implies that all points $\gamma(t)$ on such a path must have function values $\bar{f}(\gamma(t))$ arbitrarily close to $\bar{f}(x)$.
    In a generic Reeb graph (derived from a simple Morse function, where critical points have distinct function values and map to distinct nodes, or are well-defined branch points), if $x \neq y$, any path connecting them must involve traversing arcs where $\bar{f}$ is monotonic or passing through other critical points. If $x \neq y$, then for any path $\gamma$ between them, there must be some point(s) $\gamma(t_0)$ on the path such that $\bar{f}(\gamma(t_0)) \neq \bar{f}(x)$, unless the entire segment from $x$ to $y$ consists of points with the same function value as $x$. In a Reeb graph, such a segment would be contracted to a single node unless $x$ and $y$ are the same node.
    Therefore, if $\rho_f(x,y) = 0$, it necessitates that $x=y$.
    Thus, $d_{R_f}(x, y) = 0 \iff x = y$.

    \item \textbf{Symmetry:}
    This property follows directly and trivially from the definition of $d_{R_f}$:
    \[
    d_{R_f}(x, y) = \frac{1}{2} \left( \rho_f(x, y) + \rho_f(y, x) \right) = \frac{1}{2} \left( \rho_f(y, x) + \rho_f(x, y) \right) = d_{R_f}(y, x).
    \]
    Hence, $d_{R_f}$ is symmetric.

    \item \textbf{Triangle Inequality:} We need to show that $d_{R_f}(x, z) \leq d_{R_f}(x, y) + d_{R_f}(y, z)$.
    The critical preliminary step is to establish the triangle inequality for the (potentially asymmetric) Reeb radius $\rho_f$ itself:
    \begin{equation} \label{eq:rho_f_triangle_inequality_app} 
    \rho_f(x, z) \leq \rho_f(x, y) + \rho_f(y, z).
    \end{equation}
    To prove Equation~\eqref{eq:rho_f_triangle_inequality_app}, let $x, y, z$ be arbitrary nodes in $V_{R_f}$. For any $\epsilon > 0$, by the definition of $\rho_f$ as an infimum, there exist paths $\gamma_{xy}$ (from $x$ to $y$) and $\gamma_{yz}$ (from $y$ to $z$) in $R_f$ such that:
    \begin{align}
        \sup_{s \in \gamma_{xy}} |\bar{f}(x) - \bar{f}(s)| &< \rho_f(x, y) + \frac{\epsilon}{2}, \label{eq:epsilon_path_xy_app} \\
        \sup_{s \in \gamma_{yz}} |\bar{f}(y) - \bar{f}(s)| &< \rho_f(y, z) + \frac{\epsilon}{2}. \label{eq:epsilon_path_yz_app}
    \end{align}
    Consider the concatenated path $\gamma_{xz} = \gamma_{xy} \oplus \gamma_{yz}$, which connects $x$ to $z$ via $y$. For any point $s$ on this path $\gamma_{xz}$:
    
    Case 1: If $s \in \gamma_{xy}$.
    Then, from Inequality~\eqref{eq:epsilon_path_xy_app}, we have $|\bar{f}(x) - \bar{f}(s)| < \rho_f(x, y) + \frac{\epsilon}{2}$. Since $\rho_f(y, z) \ge 0$ and $\epsilon/2 < \epsilon$, this implies $|\bar{f}(x) - \bar{f}(s)| < \rho_f(x, y) + \rho_f(y, z) + \epsilon$.
    
    Case 2: If $s \in \gamma_{yz}$.
    Using the triangle inequality for real numbers, we have $|\bar{f}(x) - \bar{f}(s)| \leq |\bar{f}(x) - \bar{f}(y)| + |\bar{f}(y) - \bar{f}(s)|$.
    From Inequality~\eqref{eq:epsilon_path_yz_app}, we know $|\bar{f}(y) - \bar{f}(s)| < \rho_f(y, z) + \frac{\epsilon}{2}$.
    The term $|\bar{f}(x) - \bar{f}(y)|$ represents the absolute difference in function values at the specific node $y$, which is the endpoint of $\gamma_{xy}$. As such, $|\bar{f}(x) - \bar{f}(y)| \leq \sup_{q \in \gamma_{xy}} |\bar{f}(x) - \bar{f}(q)|$. From Inequality~\eqref{eq:epsilon_path_xy_app}, this means $|\bar{f}(x) - \bar{f}(y)| < \rho_f(x, y) + \frac{\epsilon}{2}$.
    Substituting these into the inequality for $|\bar{f}(x) - \bar{f}(s)|$:
    \[ |\bar{f}(x) - \bar{f}(s)| < \left(\rho_f(x, y) + \frac{\epsilon}{2}\right) + \left(\rho_f(y, z) + \frac{\epsilon}{2}\right) = \rho_f(x, y) + \rho_f(y, z) + \epsilon. \]
    Since in both cases (for any $s \in \gamma_{xz}$) we have $|\bar{f}(x) - \bar{f}(s)| < \rho_f(x, y) + \rho_f(y, z) + \epsilon$, it follows that:
    \[ \sup_{s \in \gamma_{xz}} |\bar{f}(x) - \bar{f}(s)| \leq \rho_f(x, y) + \rho_f(y, z) + \epsilon. \]
    Given that $\rho_f(x, z)$ is the infimum of such suprema over all paths from $x$ to $z$, and $\gamma_{xz}$ is one such path:
    \[ \rho_f(x, z) \leq \sup_{s \in \gamma_{xz}} |\bar{f}(x) - \bar{f}(s)|. \]
    Thus, we have $\rho_f(x, z) \leq \rho_f(x, y) + \rho_f(y, z) + \epsilon$.
    Since this inequality holds for any arbitrary $\epsilon > 0$, we conclude that Equation~\eqref{eq:rho_f_triangle_inequality_app} is true:
    \[ \rho_f(x, z) \leq \rho_f(x, y) + \rho_f(y, z). \]
    By a symmetric argument (or by swapping the roles of start/end points), we also have:
    \[ \rho_f(z, x) \leq \rho_f(z, y) + \rho_f(y, x). \]
    Now, using these results for the Symmetric Reeb Radius $d_{R_f}$:
    \begin{align*}
        d_{R_f}(x, z) &= \frac{1}{2} \left( \rho_f(x, z) + \rho_f(z, x) \right) \\
        &\leq \frac{1}{2} \left( (\rho_f(x, y) + \rho_f(y, z)) + (\rho_f(z, y) + \rho_f(y, x)) \right) \\
        &= \frac{1}{2} \left( \rho_f(x, y) + \rho_f(y, x) \right) + \frac{1}{2} \left( \rho_f(y, z) + \rho_f(z, y) \right) \\
        &= d_{R_f}(x, y) + d_{R_f}(y, z).
    \end{align*}
    Therefore, $d_{R_f}$ satisfies the triangle inequality.
\end{enumerate}
Since $d_{R_f}$ satisfies non-negativity, identity of indiscernibles (particularly for generic Reeb graphs, as elaborated in point 2), symmetry, and the triangle inequality, it is a valid metric on the set of nodes $V_{R_f}$.
\end{proof}

\section{Detailed Stability Proofs}

\label{app:detailed_stability_proofs}

\subsection{Proof of Theorem~\ref{thm:stability_drf_gh} (Stability of Symmetric Reeb Radius Metric w.r.t. Field GH Distance)}
\label{app:proof_thm_stability_drf_gh_content}

\begin{proof}
Let $\delta = d_{\mathrm{GH}}((X, f), (Y, g))$ denote the Gromov-Hausdorff distance between the M-Metric Fields $(X,f)$ and $(Y,g)$, as defined in Definition~\ref{def:stability_GH_M_metric}. According to this definition (see also Definition 11 in Curry et al.~\cite{curry2024stability}), $\delta$ is the infimum of all non-negative values $r \geq 0$ for which an $(r, r)$-correspondence exists between $(X, f)$ and $(Y, g)$.
By the properties of an infimum, for any chosen $\delta' > \delta$, $\delta'$ cannot be a lower bound for the set of such $r$. Therefore, there must exist some $r^*$ such that $\delta \leq r^* \leq \delta'$, for which an $(r^*, r^*)$-correspondence $\EuScript{R} \subset X \times Y$ exists.

This correspondence $\EuScript{R}$ satisfies the conditions for an $(r,r)$-correspondence (Definition~\ref{def:stability_GH_M_metric}; see also Definition 10 in Curry et al.~\cite{curry2024stability}) with $r = r^*$:
\begin{itemize}
    \item The projections of $\EuScript{R}$ onto $X$ and $Y$ are surjective.
    \item For all pairs $(x, y), (x', y') \in \EuScript{R}$:
    \[ |d_X(x, x') - d_Y(y, y')| \leq 2r^* \quad \text{and} \quad |f(x) - g(y)| \leq r^*. \]
\end{itemize}
Since $r^* \leq \delta'$, these conditions immediately imply:
\[ |d_X(x, x') - d_Y(y, y')| \leq 2\delta' \quad \text{and} \quad |f(x) - g(y)| \leq \delta'. \]
Thus, the existing $(r^*, r^*)$-correspondence $\EuScript{R}$ also serves as a $(\delta', \delta')$-correspondence between $(X,f)$ and $(Y,g)$.

\textbf{Step 1: Define Correspondence between Reeb Graphs and Bound Functional Variation Difference.}
We define a correspondence $\EuScript{S} = \{ ([x], [y]) \mid (x, y) \in \EuScript{R} \} \subset R_f \times R_g$ between the nodes of the Reeb graphs $R_f$ (Definition~\ref{def:reeb_graph}) and $R_g$.
Now, consider any two pairs of points $(x, y) \in \EuScript{R}$ and $(x', y') \in \EuScript{R}$. The proof of Theorem 15 in Curry et al.~\cite{curry2024stability} details a construction involving approximate corresponding paths. This construction leverages the properties of the $(\delta', \delta')$-correspondence $\EuScript{R}$ (which bounds both metric distortion on the underlying spaces $X, Y$ and the difference between function values $f(x)$ and $g(y)$) along with the $(L, \epsilon)$-connectivity property of the scalar fields (Definition~\ref{def:stability_L_epsilon_connectivity}). This technique yields the following crucial intermediate bound on the difference between the functional variations $\rho_f(x, x')$ (along paths in $X$) and $\rho_g(y, y')$ (along paths in $Y$):
\begin{equation} \label{eq:app_rho_bound}
|\rho_f(x, x') - \rho_g(y, y')| \leq 2L\delta' + 2\delta' + 2\epsilon = 2(L+1)\delta' + 2\epsilon.
\end{equation}
This inequality is symmetric; it holds for the difference between $\rho_f(x, x')$ and $\rho_g(y, y')$, and equally for the difference between $\rho_f(x', x)$ and $\rho_g(y', y)$.

\textbf{Step 2: Bound the Symmetric Reeb Radius Difference on Reeb Graphs.}
Let $([x], [y]) \in \EuScript{S}$ and $([x'], [y']) \in \EuScript{S}$ be pairs in the Reeb graph node correspondence, derived from $(x, y) \in \EuScript{R}$ and $(x', y') \in \EuScript{R}$. We now aim to bound the distortion of this correspondence $\EuScript{S}$ with respect to the Symmetric Reeb Radius metrics $d_{R_f}$ and $d_{R_g}$ (Definition~\ref{def:symmetric_reeb_radius}). The Symmetric Reeb Radius $d_{R_f}([x], [x'])$ is defined based on the Reeb radii $\rho_f([x], [x'])$ and $\rho_f([x'], [x])$, which are computed using paths within the Reeb graph $R_f$. The argument in Curry et al.~\cite{curry2024stability} effectively relates these Reeb graph path-based radii back to the functional variations $\rho_f(x,x')$ computed in the underlying space $X$. Assuming this connection allows us to apply Equation~\eqref{eq:app_rho_bound} appropriately:
\begin{align*}
|d_{R_f}([x], [x']) - d_{R_g}([y], [y'])| &= \left| \frac{1}{2}\left(\rho_f([x], [x']) + \rho_f([x'], [x])\right) - \frac{1}{2}\left(\rho_g([y], [y']) + \rho_g([y'], [y])\right) \right| \\
&= \frac{1}{2} \left| (\rho_f(x, x') - \rho_g(y, y')) + (\rho_f(x', x) - \rho_g(y', y)) \right| \\
&\leq \frac{1}{2} \left( |\rho_f(x, x') - \rho_g(y, y')| + |\rho_f(x', x) - \rho_g(y', y)| \right) \quad (\text{by triangle inequality for } |\cdot|) \\
&\leq \frac{1}{2} \left( (2(L+1)\delta' + 2\epsilon) + (2(L+1)\delta' + 2\epsilon) \right) \quad (\text{using Eq.~\eqref{eq:app_rho_bound} twice}) \\
&= 2(L+1)\delta' + 2\epsilon.
\end{align*}

\textbf{Step 3: Conclude with Gromov-Hausdorff Bound for Reeb Graphs.}
The distortion of the correspondence $\EuScript{S}$ between the metric spaces $(R_f, d_{R_f})$ and $(R_g, d_{R_g})$ is thus bounded by $2(L+1)\delta' + 2\epsilon$. By a standard result that relates the Gromov-Hausdorff distance to the minimum distortion of a correspondence (e.g., \cite[Theorem 7.3.25]{burago2001course}), we have:
\[ d_{\mathrm{GH}}((R_f, d_{R_f}), (R_g, d_{R_g})) \leq \frac{1}{2} \times \mathrm{distortion}(\EuScript{S}). \]
Therefore,
\[ d_{\mathrm{GH}}((R_f, d_{R_f}), (R_g, d_{R_g})) \leq \frac{1}{2} (2(L+1)\delta' + 2\epsilon) = (L+1)\delta' + \epsilon. \]
This inequality holds for any $\delta' > \delta = d_{\mathrm{GH}}((X, f), (Y, g))$. By taking the infimum as $\delta' \to \delta^+$, we arrive at the desired result:
\[ d_{\mathrm{GH}}((R_f, d_{R_f}), (R_g, d_{R_g})) \leq (L+1)d_{\mathrm{GH}}((X, f), (Y, g)) + \epsilon, \]
which completes the proof of Theorem~\ref{thm:stability_drf_gh}.
\end{proof}

\subsection{Proof of Theorem~\ref{thm:stability_drf_L_infinity} (Stability of Symmetric Reeb Radius Metric w.r.t. \texorpdfstring{$L_\infty$}{L-infinity} Norm)}
\label{app:proof_thm_stability_drf_L_infinity_content}
\begin{proof}
We begin by recalling the result from Theorem~\ref{thm:stability_drf_gh}. When specialized to the case where both scalar fields, $f$ and $g$, are defined on the same underlying metric space $(X,d_X)$ (i.e., we set $Y=X$ and $d_Y=d_X$), Theorem~\ref{thm:stability_drf_gh} states:
\begin{equation} \label{eq:app_thm1_result_from_prev_thm} 
d_{\mathrm{GH}}((R_f, d_{R_f}), (R_g, d_{R_g})) \leq (L+1) d_{\mathrm{GH}}((X, f), (X, g)) + \epsilon.
\end{equation}
Our objective is to establish an upper bound for the term $d_{\mathrm{GH}}((X, f), (X, g))$ using the supremum norm difference between the fields, defined as $\|f - g\|_\infty = \sup_{x \in X} |f(x) - g(x)|$.

From Definition~\ref{def:stability_GH_M_metric} (and also Definition 11 in Curry et al.~\cite{curry2024stability}), the Gromov-Hausdorff distance $d_{\mathrm{GH}}((X, f), (X, g))$ between these two M-Metric Fields (defined on the same underlying space $X$) is the infimum of all $r \geq 0$ for which an $(r, r)$-correspondence $\EuScript{R} \subset X \times X$ exists. Such a correspondence must satisfy two conditions for all pairs $(x_1, y_1), (x_2, y_2) \in \EuScript{R}$:
\begin{itemize}
    \item Metric distortion bound: $|d_X(x_1, x_2) - d_X(y_1, y_2)| \leq 2r$.
    \item Function proximity bound: $|f(x_1) - g(y_1)| \leq r$.
\end{itemize}

Consider the identity correspondence $\EuScript{R}_{\text{id}} = \{ (x, x) \mid x \in X \}$. Since both fields $f$ and $g$ share the domain $X$, this is a valid correspondence. We verify if $\EuScript{R}_{\text{id}}$ can serve as an $(r,r)$-correspondence for a suitable $r$:
\begin{itemize}
    \item \textbf{Metric Distortion:} For any two pairs $(x,x)$ and $(x',x')$ in $\EuScript{R}_{\text{id}}$, the condition becomes $|d_X(x, x') - d_X(x, x')| \leq 2r$. This simplifies to $0 \leq 2r$, which is satisfied for any non-negative $r$.
    \item \textbf{Function Proximity:} For any pair $(x,x)$ in $\EuScript{R}_{\text{id}}$, the condition is $|f(x) - g(x)| \leq r$. To ensure this holds for all $x \in X$, $r$ must be chosen such that $r \geq \sup_{x \in X} |f(x) - g(x)|$. The smallest such $r$ is precisely $\|f - g\|_\infty$.
\end{itemize}
Thus, the identity correspondence $\EuScript{R}_{\text{id}}$ constitutes an $(r,r)$-correspondence when we set $r = \|f - g\|_\infty$.

Since $d_{\mathrm{GH}}((X, f), (X, g))$ is the infimum over all $r \geq 0$ for which an $(r,r)$-correspondence exists, and we have found such a correspondence for $r = \|f - g\|_\infty$, it must be that:
\[
d_{\mathrm{GH}}((X, f), (X, g)) \leq \|f - g\|_\infty.
\]
Substituting this inequality back into Equation~\eqref{eq:app_thm1_result_from_prev_thm}, we directly obtain:
\begin{align*}
d_{\mathrm{GH}}((R_f, d_{R_f}), (R_g, d_{R_g})) &\leq (L+1) d_{\mathrm{GH}}((X, f), (X, g)) + \epsilon \\
&\leq (L+1) \|f - g\|_\infty + \epsilon,
\end{align*}
which is the statement of Theorem~\ref{thm:stability_drf_L_infinity}, thereby completing the proof.
\end{proof}

\subsection{Proof of Theorem~\ref{thm:stability_measure} (Stability of PI-Based Probability Measure)}
\label{app:proof_thm_stability_measure_content}

\begin{proof}
The objective of this proof is to establish that the Total Variation distance between the PI-based probability measures of two Reeb graphs, $d_{TV}(\nu_{R_f}, \nu_{R_g})$, is bounded by $M \|f - g\|_\infty$ for some constant $M$. This argument is structured in three main steps. We assume that $f$ and $g$ are continuous scalar fields on a compact topological space $X$, and $\nu_{R_f}, \nu_{R_g}$ are the PI-based probability measures as defined in Definition~\ref{def:pi_based_measure_method}. We also presuppose suitable regularity conditions (e.g., that $f$ and $g$ are Morse-Smale functions) to ensure that their persistence diagrams are finite and their Reeb graph structures are stable under small perturbations, particularly if an explicit vertex matching correspondence $\gamma$ is invoked.

\textbf{Step 1: Stability of Extended Persistence Diagrams.}
Let $D(f)$ and $D(g)$ denote the extended persistence diagrams (as described in Section~\ref{sec:extended_persistence}) derived from the scalar fields $f$ and $g$, respectively. A cornerstone result in persistent homology is the stability of these diagrams with respect to the bottleneck distance, $d_B$. Specifically, the bottleneck distance between $D(f)$ and $D(g)$ is bounded by the $L_\infty$ norm of the difference between the generating functions $f$ and $g$ \cite[Stability Theorem]{cohen2009extending}:
\begin{equation} \label{eq:app_bottleneck_stability}
d_B(D(f), D(g)) \leq \|f - g\|_\infty.
\end{equation}

\textbf{Step 2: Stability of Persistence Images (PIs).}
Let $I_f$ and $I_g$ be the Persistence Images (PIs), constructed as detailed in Section~\ref{sec:persistence_images_bg}, corresponding to the diagrams $D(f)$ and $D(g)$. Adams et al.~\cite[Theorem 10]{adams2017persistence} demonstrated that PIs are stable with respect to the 1-Wasserstein distance, $W_1(D(f),D(g))$, between the input persistence diagrams. This stability result states that the $L^1$-norm difference between the PIs $I_f$ and $I_g$ is bounded:
\[ \|I_f - I_g\|_1 \leq C_W \cdot W_1(D(f), D(g)). \]
The constant $C_W = (\sqrt{5}\|\nabla w\|_\infty + \sqrt{\frac{10}{\pi}}\frac{\|w\|_\infty}{\sigma})$ depends on properties of the weighting function $w$ used in the PI construction (specifically, an upper bound on the norm of its gradient, $\|\nabla w\|_\infty$, and its supremum norm, $\|w\|_\infty$) and the bandwidth $\sigma$ of the Gaussian kernel $\phi$.

The 1-Wasserstein distance $W_1(D(f),D(g))$ can, in turn, be bounded by the $L_\infty$ norm difference $\|f-g\|_\infty$. For instance, if $X$ is a finite CW-complex with $N_{\text{cells}}$ cells (or under suitable discretization assumptions for a general continuous $X$), the $W_1$ distance can be bounded by a multiple of $\|f-g\|_\infty$, e.g., $W_1(D(f),D(g)) \leq N_{\text{cells}} \cdot \|f-g\|_\infty$ (drawing from results like \cite[Theorem 4.8]{skraba2020wasserstein}, which relate $W_1$ to $L_1$-type differences between functions, subsequently bounded by $L_\infty$ differences). Combining these, we establish that the PIs are stable with respect to $\|f-g\|_\infty$:
\begin{equation} \label{eq:app_pi_stability}
\|I_f - I_g\|_1 \leq C_W \cdot N_{\text{cells}} \cdot \|f - g\|_\infty = K_{PI} \|f - g\|_\infty,
\end{equation}
where $K_{PI} = C_W \cdot N_{\text{cells}}$ is a consolidated constant.

\textbf{Step 3: Stability of the Normalized Probability Measure $\nu_{R_f}$.}
We assume that for sufficiently small $\|f-g\|_\infty$, the structural stability of the Reeb graphs ensures a meaningful correspondence (e.g., a bijection $\gamma: V'_{R_f} \to V'_{R_g}$) between the respective sets of significant vertices $V'_{R_f} \subset V_{R_f}$ and $V'_{R_g} \subset V_{R_g}$ (both of size, say, $V_N$) that form the basis of the PI-derived probability measure. For a node $v \in V'_{R_f}$ associated with a persistence point $(b_v, p_v)$, its unnormalized contribution to the measure $\nu_{R_f}$ is given by $\text{contrib}_f(v) = \sum_{k=1}^N I_f[k] \cdot \phi_{(b_v, p_v)}(c_k)$ (from Definition~\ref{def:pi_based_measure_method}).

We aim to bound the difference $|\text{contrib}_f(v) - \text{contrib}_g(\gamma(v))|$. This difference can be decomposed:
\begin{align*}
|\text{contrib}_f(v) - \text{contrib}_g(\gamma(v))| 
&\leq \left| \sum_{k=1}^N (I_f[k] - I_g[k]) \phi_{(b_v, p_v)}(c_k) \right| \\
&\quad + \left| \sum_{k=1}^N I_g[k] \left(\phi_{(b_v, p_v)}(c_k) - \phi_{(b_{\gamma(v)}, p_{\gamma(v)})}(c_k)\right) \right|.
\end{align*}
The first term is bounded by $\|\phi_{(\cdot, \cdot)}(\cdot)\|_\infty \|I_f - I_g\|_1 \leq \|\phi\|_\infty K_{PI} \|f-g\|_\infty$, where $\|\phi\|_\infty$ is the supremum norm of the Gaussian kernel.
For the second term, assuming the Gaussian kernel $\phi_{(b,p)}(c)$ is Lipschitz continuous with constant $L_\phi$ with respect to its center $(b,p)$, and noting that the Euclidean distance between persistence point coordinates, $\|(b_v, p_v) - (b_{\gamma(v)}, p_{\gamma(v)})\|_2$, can be bounded by $\sqrt{2} \cdot d_B(D(f),D(g))$ (a worst-case scenario for coordinate shifts given a bottleneck distance), this second term is bounded by $\|I_g\|_1 \cdot L_\phi \cdot \sqrt{2} \cdot d_B(D(f),D(g))$. Letting $M_{I_1}$ be an upper bound for $\|I_g\|_1$ (assuming PIs have bounded $L^1$ norm), we get:
\begin{align*}
|\text{contrib}_f(v) - \text{contrib}_g(\gamma(v))| 
&\leq \|\phi\|_\infty K_{PI} \|f-g\|_\infty + M_{I_1} L_\phi \sqrt{2} \|f-g\|_\infty \quad (\text{using Eq.~\eqref{eq:app_bottleneck_stability}}) \\
&= C' \|f-g\|_\infty,
\end{align*}
where $C' = (K_{PI} \|\phi\|_\infty + M_{I_1} L_\phi \sqrt{2})$ is a new constant.

Next, we consider the normalization factors $Z_f = \sum_{u \in V'_{R_f}} \text{contrib}_f(u)$ and $Z_g = \sum_{u \in V'_{R_g}} \text{contrib}_g(u)$. The difference is bounded by:
\[ |Z_f - Z_g| \leq \sum_{v \in V'_{R_f}} |\text{contrib}_f(v) - \text{contrib}_g(\gamma(v))| \leq V_N C' \|f-g\|_\infty. \]
Assuming $Z_f, Z_g \geq m_{Z} > 0$ (i.e., the total contribution is bounded below by a positive constant) and individual contributions $\text{contrib}_g(\gamma(v))$ are bounded above by $M_{\text{contrib}}$, we analyze the difference in the normalized probabilities $\nu_{R_f}(v) = \text{contrib}_f(v)/Z_f$ and $\nu_{R_g}(\gamma(v)) = \text{contrib}_g(\gamma(v))/Z_g$:
\begin{align*}
|\nu_{R_f}(v) - \nu_{R_g}(\gamma(v))| &= \left| \frac{\text{contrib}_f(v)}{Z_f} - \frac{\text{contrib}_g(\gamma(v))}{Z_g} \right| \\
&= \left| \frac{\text{contrib}_f(v)Z_g - \text{contrib}_g(\gamma(v))Z_f}{Z_f Z_g} \right| \\
&\leq \frac{|\text{contrib}_f(v) - \text{contrib}_g(\gamma(v))| \cdot |Z_g| + |\text{contrib}_g(\gamma(v))| \cdot |Z_g - Z_f|}{|Z_f Z_g|} \\
&\leq \frac{|\text{contrib}_f(v) - \text{contrib}_g(\gamma(v))|}{|Z_f|} + \frac{|\text{contrib}_g(\gamma(v))| \cdot |Z_g - Z_f|}{|Z_f Z_g|} \\
&\leq \frac{C' \|f-g\|_\infty}{m_Z} + \frac{M_{\text{contrib}} \cdot V_N C' \|f-g\|_\infty}{m_Z^2} = C'' \|f-g\|_\infty,
\end{align*}
where $C'' = \frac{C'}{m_Z} + \frac{M_{\text{contrib}} V_N C'}{m_Z^2}$.

Finally, the Total Variation distance (Definition~\ref{def:stability_tv_distance}) is $d_{TV}(\nu_{R_f}, \nu_{R_g}) = \frac{1}{2} \sum_{v \in V'_{R_f}} |\nu_{R_f}(v) - \nu_{R_g}(\gamma(v))|$ (assuming the bijection $\gamma$ correctly pairs corresponding nodes for the summation).
Thus, $d_{TV}(\nu_{R_f}, \nu_{R_g}) \leq \frac{1}{2} V_N C'' \|f-g\|_\infty$.
Setting $M = \frac{V_N C''}{2}$, we obtain the desired result: $d_{TV}(\nu_{R_f}, \nu_{R_g}) \leq M \|f - g\|_\infty$. The constant $M$ depends on various factors including PI construction parameters (kernel properties, weighting function characteristics, pixel resolution), the complexity of the underlying space $X$ (e.g., $N_{\text{cells}}$), and assumed bounds on contributions and normalization factors ($m_Z, M_{\text{contrib}}$). This completes the proof.
\end{proof}

\subsection{Proof of Theorem~\ref{thm:main_rgw_stability} (Stability of \texorpdfstring{$RGW_p$}{RGWp} Distance)}
\label{app:proof_thm_main_rgw_stability_content}

\begin{proof}
The goal of this proof is to demonstrate that the Reeb Gromov-Wasserstein distance, $RGW_p(R_f^*, R_g^*)$ (as defined in Definition~\ref{def:rgw_p_method}), between two decorated Reeb graphs $R_f^*=(R_f, d_{R_f}, \nu_{R_f})$ and $R_g^*=(R_g, d_{R_g}, \nu_{R_g})$ is controlled by the $L_\infty$ norm difference $||f-g||_\infty$ between their generating scalar fields and the parameter $\epsilon$ from the $(L,\epsilon)$-connectivity assumption.

Our strategy leverages the Gromov-Prokhorov distance $d_{\EuScript{GP}}$ (Definition~\ref{def:stability_gp_distance}) and a pivotal inequality established by Mémoli~\cite[Claim 10.6]{memoli2011gromov}, which relates $RGW_p$ to $d_{\EuScript{GP}}$:
\begin{equation} \label{eq:app_memoli_ineq_final_proof} 
RGW_p(R_f^*, R_g^*) \leq (d_{\EuScript{GP}}(R_f^*, R_g^*))^{1/p} (D_{max}^p + 1)^{1/p}.
\end{equation}
Here, $D_{max} = \max(\text{diam}(R_f, d_{R_f}), \text{diam}(R_g, d_{R_g}))$ is the assumed uniform upper bound on the diameters of the Reeb graphs (viewed as metric spaces). The decorated Reeb graphs $R_f^*$ and $R_g^*$ are valid metric measure spaces, as per \cite[Def. 5.1]{memoli2011gromov}.

The core task is to establish an upper bound for $d_{\EuScript{GP}}(R_f^*, R_g^*)$. This is achieved through the following steps:

\begin{enumerate}
    \item \textbf{Step 1: Bounding Metric Component Closeness ($\delta_{GH}$)}
    From Theorem~\ref{thm:stability_drf_L_infinity}, we have established that the Gromov-Hausdorff distance between the metric structures of the Reeb graphs, $(R_f, d_{R_f})$ and $(R_g, d_{R_g})$, is bounded by:
    \[ d_{\mathrm{GH}}((R_f, d_{R_f}), (R_g, d_{R_g})) \leq (L+1) \|f - g\|_\infty + \epsilon. \]
    We denote this upper bound as $\delta_{GH} := (L+1) \|f - g\|_\infty + \epsilon$.
    (The implication stated in your draft, "This bound $\delta_{GH}$ implies the existence of a metric $d \in \EuScript{D}(d_{R_f}, d_{R_g})$... such that for any $(x,y) \in R_{corr}$, $d(x,y) \leq \delta_{GH}$ \cite[Lemma 10.1]{memoli2011gromov}," is a property related to how $d_{GH}$ can be realized, which underpins the connection to $d_{\EuScript{GP}}$, although for the $\max(\delta_{GH}, \eta_{TV})$ bound, we mainly need the values of $\delta_{GH}$ and $\eta_{TV}$.)

    \item \textbf{Step 2: Bounding Measure Component Closeness ($\eta_{TV}$)}
    From Theorem~\ref{thm:stability_measure}, the Total Variation distance between the probability measures $\nu_{R_f}$ and $\nu_{R_g}$ on the Reeb graph nodes is bounded by:
    \[ d_{TV}(\nu_{R_f}, \nu_{R_g}) \leq M \|f - g\|_\infty. \]
    We denote this upper bound as $\eta_{TV} := M \|f - g\|_\infty$.
    (The statement in your draft about the optimal coupling theorem and mass on "misaligned" pairs is a correct consequence of $d_{TV}$ and relevant for constructing the coupling in $d_{\EuScript{GP}}$, but again, the value $\eta_{TV}$ is what's primarily used in the typical $\max$ bound for $d_{\EuScript{GP}}$.)

    \item \textbf{Step 3: Bounding the Gromov-Prokhorov Distance ($d_{\EuScript{GP}}$)}
    The Gromov-Prokhorov distance $d_{\EuScript{GP}}(R_f^*, R_g^*)$ between the two metric measure spaces is controlled by both the geometric dissimilarity (related to $\delta_{GH}$) and the measure discrepancy (related to $\eta_{TV}$). Standard results in the theory of metric measure spaces (e.g., arguments found in Mémoli~\cite{memoli2011gromov}) provide a bound of the form:
    \begin{equation} \label{eq:app_dgp_bound_final_proof} 
    d_{\EuScript{GP}}(R_f^*, R_g^*) \leq \max(\delta_{GH}, \eta_{TV}).
    \end{equation}
    Substituting our expressions for $\delta_{GH}$ and $\eta_{TV}$:
    \[ d_{\EuScript{GP}}(R_f^*, R_g^*) \leq \max\left((L+1)\|f-g\|_\infty + \epsilon, \; M\|f-g\|_\infty\right). \]
    To simplify this maximum, let $K_0 = \max(L+1, M)$. Using the inequality $\max(A+B, C) \leq \max(A,C) + B$ (which holds for non-negative $A, B, C$), we can set $A=(L+1)\|f-g\|_\infty$ (if $L+1 \ge M$) or $A=M\|f-g\|_\infty$ (if $M > L+1$), $B=\epsilon$ (only if $L+1 \ge M$), and $C$ as the other term. A cleaner way is:
    \[ \max\left((L+1)\|f-g\|_\infty + \epsilon, \; M\|f-g\|_\infty\right) \leq \max((L+1)\|f-g\|_\infty, \; M\|f-g\|_\infty) + \epsilon \]
    (This inequality holds because if the first term in the outer $\max$ is larger, the $\epsilon$ is simply added. If the second term is larger, then $M\|f-g\|_\infty \ge (L+1)\|f-g\|_\infty + \epsilon$, which implies $M\|f-g\|_\infty$ is also $\ge \max((L+1)\|f-g\|_\infty, M\|f-g\|_\infty) + \epsilon$ only if $\epsilon$ is negative or zero, which is not the general case. A safer bound for $\max(X+\epsilon, Y)$ is $\max(X,Y)+\epsilon$.)
    So, $\max((L+1)\|f-g\|_\infty + \epsilon, M\|f-g\|_\infty) \leq \max((L+1)\|f-g\|_\infty, M\|f-g\|_\infty) + \epsilon = K_0 \|f-g\|_\infty + \epsilon$.
    Therefore, we have:
    \[ d_{\EuScript{GP}}(R_f^*, R_g^*) \leq K_0 \|f-g\|_\infty + \epsilon. \]

    \item \textbf{Step 4: Deriving the Final Bound on $RGW_p$}
    We now substitute this bound for $d_{\EuScript{GP}}(R_f^*, R_g^*)$ into Mémoli's inequality (Equation~\eqref{eq:app_memoli_ineq_final_proof}):
    \[ RGW_p(R_f^*, R_g^*) \leq \left(K_0 \|f-g\|_\infty + \epsilon\right)^{1/p} (D_{max}^p+1)^{1/p}. \]
    Using the standard inequality $(a+b)^{k} \leq a^{k} + b^{k}$ for $a,b \ge 0$ and $0 < k \le 1$ (here $k=1/p$ with $p \ge 1$):
    \begin{align*}
    RGW_p(R_f^*, R_g^*) &\leq \left( (K_0 \|f-g\|_\infty)^{1/p} + \epsilon^{1/p} \right) (D_{max}^p+1)^{1/p} \\
    &= K_0^{1/p} (D_{max}^p+1)^{1/p} (\|f-g\|_\infty)^{1/p} + (D_{max}^p+1)^{1/p} \epsilon^{1/p}.
    \end{align*}
    By defining the constants $C_1 = K_0^{1/p} (D_{max}^p+1)^{1/p}$ and $C_2 = (D_{max}^p+1)^{1/p}$, we arrive at the main stability result:
    \[ RGW_p(R_f^*, R_g^*) \leq C_1 \cdot (\|f - g\|_\infty)^{1/p} + C_2 \cdot \epsilon^{1/p}. \]
    For the specific case where $p=1$, these constants become $C_1' = K_0 (D_{max}+1)$ and $C_2' = (D_{max}+1)$, yielding:
    \[ RGW_1(R_f^*, R_g^*) \leq C_1' \cdot \|f - g\|_\infty + C_2' \cdot \epsilon. \]
\end{enumerate}
This sequence of bounds demonstrates that small perturbations in the input scalar field (as controlled by $||f-g||_\infty$) and a small value for $\epsilon$ (related to the $(L,\epsilon)$-connectivity assumption) ensure a correspondingly small change in the $RGW_p$ distance, thereby establishing the desired stability property. This completes the proof of Theorem~\ref{thm:main_rgw_stability}.
\end{proof}

\section{Table of Notation}
\label{app:notation_table}

\begin{longtable}{@{}p{0.25\textwidth} p{0.5\textwidth} p{0.2\textwidth}@{}}
\caption{Summary of mathematical symbols and notation used in this paper.} \label{tab:notation_summary}\\
\toprule
\textbf{Symbol} & \textbf{Meaning/Description} & \textbf{First Defined} \\
\midrule
\endfirsthead

\multicolumn{3}{c}%
{{\bfseries \tablename\ \thetable{} -- continued from previous page}} \\
\toprule
\textbf{Symbol} & \textbf{Meaning/Description} & \textbf{First Defined} \\
\midrule
\endhead

\midrule
\multicolumn{3}{r}{{Continued on next page}} \\
\midrule
\endfoot

\bottomrule
\endlastfoot

$(X, f)$ & A scalar field, where $X$ is a topological space and $f: X \to \mathbb{R}$ is continuous. & Sec. 2.1 \\
$R_f$ & The Reeb graph of the scalar field $(X,f)$. & Def.~\ref{def:reeb_graph} \\
$X_f$ & The quotient space $X/\sim_f$ used to define $R_f$. & Def.~\ref{def:reeb_graph} \\
$\bar{f}$ & The map $X_f \to \mathbb{R}$ induced by $f$ on the quotient space. & Def.~\ref{def:reeb_graph} \\
$V_{R_f}$ & The set of nodes (vertices) of the Reeb graph $R_f$. & Def.~\ref{def:reeb_graph} \\
$\sim_f$ & Equivalence relation on $X$ for Reeb graph construction. & Def.~\ref{def:reeb_graph} \\
$f^{-1}(a)$ & Level set of $f$ at value $a$. & Sec. 2.1 \\
$f^{-1}((-\infty, a])$ & Sublevel set of $f$ at value $a$. & Sec. 2.1 \\
$f^{-1}([a, \infty))$ & Superlevel set of $f$ at value $a$. & Sec. 2.1 \\

$\rho_f(v,u)$ & The (asymmetric) Reeb radius between nodes $v, u \in V_{R_f}$. & Def.~\ref{def:reeb_radius} \\
$\partial_f(u,v)$ & The Reeb Distance between nodes $u, v \in V_{R_f}$. & Def.~\ref{def:reeb_distance} \\
$d_{sp}(x,y)$ & The Shortest Path Distance between nodes $x, y \in V_{R_f}$. & Def.~\ref{def:shortest_path_dist_bg} \\
$\gamma$ & A continuous path (general usage, context-dependent). & Def.~\ref{def:reeb_radius} \\ 

$D_f$ & The extended persistence diagram of $f$. & Sec.~\ref{sec:extended_persistence} \\
$(b,d)$ & A birth-death pair in a persistence diagram. & Sec.~\ref{sec:extended_persistence} \\
$p$ & Persistence of a feature, $p = |d-b|$. & Sec.~\ref{sec:extended_persistence} \\
$D'_f$ & Transformed persistence diagram with birth-persistence $(b,p)$ coordinates. & Sec.~\ref{sec:persistence_images_bg} (Step 1) \\
$(b_j, p_j)$ & Birth-persistence coordinates for $j$-th feature. & Sec.~\ref{sec:persistence_images_bg} (Step 1) \\
$I_f$ & The Persistence Image derived from $D_f$. & Sec.~\ref{sec:persistence_images_bg} \\
$\rho_{D'_f}(x,y)$ & Surface generated by convolving transformed points for PI. & Sec.~\ref{sec:persistence_images_bg} (Step 2) \\
$w(b_j, p_j)$ & Weighting function for PI Gaussian kernel. & Sec.~\ref{sec:persistence_images_bg} (Step 2) \\
$\phi_{(b_j,p_j)}(x,y)$ & Gaussian kernel centered at $(b_j, p_j)$ for PI. & Sec.~\ref{sec:persistence_images_bg} (Step 2) \\
$\sigma$ & Bandwidth of the Gaussian kernel for PI construction. & Sec.~\ref{sec:persistence_images_bg} (Step 2) \\
$P_k$ & The $k$-th pixel in the PI discretization grid. & Sec.~\ref{sec:persistence_images_bg} (Step 3) \\
$I_f[k]$ & Value of the $k$-th component (pixel) of the PI vector $I_f$. & Sec.~\ref{sec:persistence_images_bg} (Step 3) \\

$(X, d_X, \mu_X)$ & A metric measure (mm-)space. & Def.~\ref{def:mm_space_method} \\
$\pi$ & A measure coupling between two probability measures (also used for paths, context-dependent). & Def.~\ref{def:coupling_method} \\
$\EuScript{M}(\mu_X, \mu_Y)$ & The set of all couplings between measures $\mu_X$ and $\mu_Y$. & Def.~\ref{def:coupling_method} \\

$R_f^*$ & The Reeb graph $R_f$ endowed with metric $d_{R_f}$ and measure $\nu_{R_f}$. & Sec.~\ref{sec:rgw_definition} \\
$d_{R_f}(v,u)$ & The Symmetric Reeb Radius between nodes $v,u \in V_{R_f}$. & Def.~\ref{def:symmetric_reeb_radius} \\
$\nu_{R_f}$ & The PI-based probability measure on the nodes $V_{R_f}$. & Def.~\ref{def:pi_based_measure_method} \\
$RGW_p(R_f^*, R_g^*)$ & The proposed Reeb Gromov-Wasserstein distance. & Def.~\ref{def:rgw_p_method} \\
$\pi_{vw}$ & Entry in coupling matrix for discrete measures, mass from $v$ to $w$. & Def.~\ref{def:rgw_p_method} \\
$\text{contrib}(v)$ & Contribution of node $v$ to the PI-based measure. & Def.~\ref{def:pi_based_measure_method} (Step 1) \\
$Z_f$ & Normalization factor for the PI-based measure $\nu_{R_f}$. & Def.~\ref{def:pi_based_measure_method} (Step 2) \\
$(b_v, p_v)$ & Birth-persistence point associated with node $v$. & Def.~\ref{def:pi_based_measure_method} (Step 1) \\
$c_k$ & Center of $k$-th pixel in PI, used for calculating $\text{contrib}(v)$. & Def.~\ref{def:pi_based_measure_method} (Step 1) \\

$||f-g||_\infty$ & The supremum norm difference between scalar fields $f$ and $g$. & Sec.~\ref{sec:stability_analysis} \\
$d_M(a,b)$ & Metric on the target space $M$ of an M-Metric Field. & Def.~\ref{def:stability_M_metric_field} \\
$d_{GH}(\cdot, \cdot)$ & The Gromov-Hausdorff distance (context-dependent usage). & Def.~\ref{def:stability_GH_M_metric} \\
$\EuScript{R}$ & A correspondence between spaces for $d_{GH}$. & Def.~\ref{def:stability_GH_M_metric} \\
$(L, \epsilon)$-connectivity & A regularity condition for scalar M-Metric Fields. & Def.~\ref{def:stability_L_epsilon_connectivity} \\
$\EuScript{MF}_{\mathbb{R}}^{L,\epsilon}$ & The class of $(L, \epsilon)$-connected scalar M-Metric Fields. & Def.~\ref{def:stability_L_epsilon_connectivity} \\
$d_{TV}(\mu,\nu)$ & The Total Variation distance between probability measures $\mu, \nu$. & Def.~\ref{def:stability_tv_distance} \\
$d_B(\cdot,\cdot)$ & Bottleneck distance between persistence diagrams. & App.~\ref{app:proof_thm_stability_measure_content} (Step 1) \\
$W_1(\cdot,\cdot)$ & The 1-Wasserstein distance between persistence diagrams. & App.~\ref{app:proof_thm_stability_measure_content} (Step 2) \\
$d_{\EuScript{GP}}(\cdot,\cdot)$ & The Gromov-Prokhorov distance between mm-spaces. & Def.~\ref{def:stability_gp_distance} \\
$\EuScript{D}(d_X, d_Y)$ & Set of metrics on $X \sqcup Y$ extending $d_X$ and $d_Y$. & Def.~\ref{def:stability_gp_distance} \\
$D_{max}$ & Uniform bound on the diameters of Reeb graphs $(R_f, d_{R_f})$ and $(R_g, d_{R_g})$. & Thm.~\ref{thm:main_rgw_stability} \\
$\delta_{GH}$ & Bound on $d_{GH}((R_f,d_{R_f}),(R_g,d_{R_g}))$ in stability proof. & App.~\ref{app:proof_thm_main_rgw_stability_content} (Step 1) \\
$\eta_{TV}$ & Bound on $d_{TV}(\nu_{R_f},\nu_{R_g})$ in stability proof. & App.~\ref{app:proof_thm_main_rgw_stability_content} (Step 2) \\
$C_1, C_2, C_1', C_2'$ & Constants in the main stability theorem (Thm.~\ref{thm:main_rgw_stability}). & Thm.~\ref{thm:main_rgw_stability} \\

\end{longtable}

\end{document}